\documentclass[11pt, draftcls, onecolumn, letterpaper]{IEEEtran}
\usepackage{color,overpic,cite,pdfsync,url}
\usepackage{amsmath,amsfonts,amssymb,amsthm,bbm,bm}
\usepackage{epstopdf}
\usepackage{fancyhdr}
\fancypagestyle{firstpage}{
  \fancyhf{}
  \fancyhead[C]{\textbf{A version of this paper has been accepted for publication in IEEE INFOCOM 2016.}}
  \fancyfoot[C]{-~\thepage~-}
}

\newcommand{\mean}[1]{\mathbb{E}\!\left[#1\right]}

\newcommand{\argmin}{\arg\min}
\newcommand{\argmax}{\arg\max}

\newtheorem{theorem}{Theorem}
\newtheorem{lemma}{Lemma}

\def\ind{\mathbbm{1}}

\def\PP{\mathbb{P}}
\def\EE{\mathbb{E}}
\title{Streaming Big Data meets Backpressure in Distributed Network Computation 
}
\author{
	Apostolos Destounis$^*$, Georgios S. Paschos$^*$, and Iordanis Koutsopoulos$^{\dagger}$\\ $^*$Huawei Technologies France Research Center \\ $^{\dagger}$Athens University of Economics and Business
	}
\begin{document}
	\maketitle
	\thispagestyle{firstpage}
	\begin{abstract}
		We study network response to queries that require  computation of remotely located data and seek to characterize the performance limits in terms of maximum sustainable query rate that can be satisfied.
		The available resources include
		(\textit{i}) a communication network graph with links over which data is routed,
		(\textit{ii}) computation nodes, over which computation load is balanced, and
		(\textit{iii}) network nodes that need to schedule raw and processed data transmissions.
		Our aim is to design a universal methodology and distributed algorithm to adaptively allocate resources in order to support \emph{maximum query rate}.
		The proposed algorithms extend in a nontrivial way the backpressure (BP) algorithm to take into account computations operated over query streams. They contribute to the fundamental understanding of network computation performance limits when the query rate is limited by both the communication bandwidth and the computation capacity, a classical setting that arises in streaming big data applications in network clouds and fogs.
	\end{abstract}
	\vspace{-0.05in}
	\section{Introduction} \label{sec:1}
	
	In recent years, the gamut of services and applications that rely on big data analytics and computations has significantly expanded. The game-changer in these platforms lies in their ability to perform computations and deliver results in real time, in the form of a service or an application. This proliferation is much attributed to the advent of smart-phones and wearable devices with multi-modal embedded sensors that facilitate data collection, and it has created the need for impromptu service delivery to the mobile user. For instance, mobile augmented-reality apps rely on real-time data retrieval from distributed data sources to offer a sense of an augmented world, supplemental to the real one.
	In mobile crowd-sensing apps, smart-phones contribute data which is aggregated, and the aggregate is provided in real time as a service to app subscribers. Further, the mobile health sector supports real-time personalized medical advice 
	based on analytics on dynamic diverse data collected from smart-phones and wearable devices to help people self-manage their health.
	
	Data computations and analytics may be performed either (\textit{i}) at the back-end i.e. at large-scale computation platforms or high-performance computing clouds of interconnected nodes with computation and storage capabilities, or (\textit{ii}) at network-edge components i.e. mobile devices or base stations, according to the newly coined concept of edge computing and the ``fog'' \cite{Bonomi12}  wherein nodes with computation and storage resources are wirelessly connected.
	
	A unifying model that captures the scenarios above is the following. A set of nodes are connected in a network through links of certain communication bandwidth, and each node has some computation capacity resources. Sequences of queries for computation are generated in a streaming fashion. Each query sequence is characterized by a type of computation, the sources where the data are collected from, and the destination where results are to be delivered. 
	In order to satisfy each computation request, an algorithm is needed to perform the following tasks: (\textit{i}) first, \textit{retrieval} of data pertinent to the query, possibly from multiple source nodes in the network. These may be either nodes that hold stored data such as databases in a computing cluster or mobile devices that provide data on the spot. (\textit{ii}) Next, \textit{determination of computation nodes} in the network that will do computations on the data; these nodes may have diverse computation resources. 
	Computation may involve aggregates, functions of or statistics on the data. (\textit{iii}) \textit{Multi-hop routing} of the unprocessed (raw) data through the network from the source nodes to computation nodes, and \textit{multi-hop routing} of the computation results (processed data) from computation nodes to the destination, (\textit{iv}) \textit{scheduling} of traffic streams of unprocessed and processed data corresponding to different queries through computation nodes of limited computational capacity and through links of limited bandwidth.
	
	\begin{figure}[t!]
		\begin{center}
			\begin{overpic}[scale=0.85]{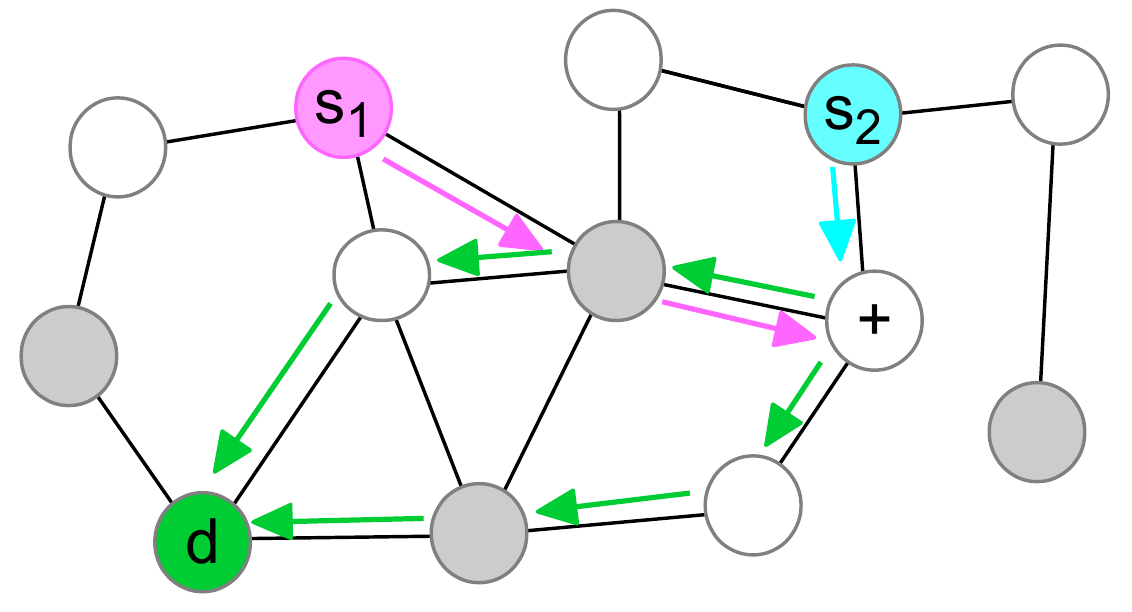}
				\put(15.5,50){\small \textbf{data source 1}}
				\put(61,50){\small \textbf{data source 2}}
				\put(1,-3){\small \textbf{destination for query response}}
				\put(83,27){\small \textbf{computation node}}
			\end{overpic}
			\caption{Illustration of network computations. Shaded nodes are forwarding ones, i.e. without computation capabilities, and white nodes have computation capabilities. Colorful nodes are sources and destinations, and colorful arrows denote routing of raw and processed data.} 
			\vspace{-0.4in}
			\label{fig:network}
		\end{center}
	\end{figure}
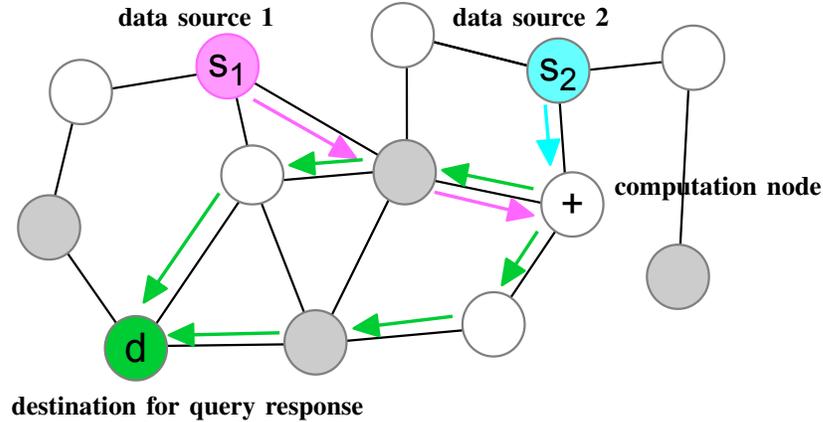
	In this work, we ask the following question: Given a network graph $\mathcal{G} = (\mathcal{N},\mathcal{L})$ with links of limited communication bandwidth and nodes of limited computation resources, what are the \textit{performance limits} of in-network computation throughput? Namely, what is the \textit{maximum rate} with which computation results can be conveyed to the destination when computations take place in the network? This question is a fundamental one to resolve in order to efficiently handle the large volume of data analytics requests by optimally utilizing system resources.
	
	\subsection{Motivating Example}
	
	Consider a simple query involving three nodes (two sources $1,2$ and destination $d$), which form a fully connected undirected graph. Let the node set be $\{1,2,d\}$ and let the link set be $\{(1,2), (1,d), (2,d)\}$. Let $C_i$ be the available computation capacity of node $i \in \{1,2,d\}$, measured in number of processed packets per second and $R_{ij}$ be the available communication bandwidth of link $(i,j)$, in packets/sec. Let $x_i$, $i=1,2$ be a datum of source $i$, $i=1,2$. Consider a stream of queries with rate $\lambda$ queries/sec where each query seeks to compute, say the sum of a datum of source $1$ and a datum of source $2$ and deliver the result to $d$.
	
	If we restrict ourselves to single-path routing, the query stream can be handled in three different ways:
	\begin{enumerate}
		\item Source $1$ sends data $x_1$ to source $2$ over link bandwidth $R_{12}$. This leads to incoming data rate $\min\{\lambda,R_{12}\}$ to source $2$. Source $2$ performs addition with its own data $x_2$ (of rate $\lambda$) and generates sums $x_1 + x_2$ at rate $\min\{C_2, \lambda, R_{12}\}$. It then sends the sums to the destination $d$ over link bandwidth $R_{2d}$. Here the computation is performed at node 2, and the rate with which sums are received at $d$ is $\min\{ C_2,\lambda, R_{12},R_{2d} \}$.
		\item Source $2$ sends data $x_2$ to source $1$ over link bandwidth $R_{12}$. Source $1$ performs addition with its data $x_1$ (of rate $\lambda$), it generates sums $x_1 + x_2$ and sends them to the destination $d$ over link bandwidth $R_{1d}$. Here, the computation is performed at node $1$, and the rate with which sums are received at $d$ is $\min\{ C_1, \lambda, R_{12},R_{1d} \}$.
		\item Source $1$ sends data $x_1$ to $d$ over link of bandwidth $R_{1d}$, and source $2$ sends data $x_2$ to $d$ over link bandwidth $R_{2d}$. The destination $d$ performs the addition and generates sums $x_1 + x_2$. Thus, the computation is performed at $d$, and the rate with which the sums are generated is $\min\{ C_d,\lambda, R_{1d},R_{2d}\}$.
	\end{enumerate}
	
	Clearly, the maximum rate of received sums depends on which computation node was used, and the routing of data on the network. The problem becomes further complicated if we allow routing through multiple paths. Moreover, considering a stream of similar queries, it is possible to load balance queries over the different options, and hence the problem obtains a multi-commodity form. The static scenario described above serves as a prelude to the dynamic problem that arises in the presence of unknown dynamic query arrivals and accumulated traffic loads at various queues in the network.
	The decisions in the dynamic scenario concern determination of the node to perform the computation for each query, as well as queue management through traffic routing, link bandwidth sharing and computation capacity allocation, and must be made adaptively.
	
	\subsection{Related Work}
	
	The problem of in-network computation has attracted a lot of attention recently. If network coding is allowed, cut-set bounds for the computational capacity of networks defined on Directed Acyclic Graphs (DAGs) are proven in \cite{Appuswami11}. These cut-set bounds cannot be achieved by routing alone, and proper network codes need to be used. However, in this paper we restrict ourselves to routing-only policies, which simplify adaptivity and distributed implementation. Prior works pertaining to routing-only approaches study the problem in a static setup, with network flows as variables \cite{Shah11}. The problem of finding an optimal flow when there is a computation cost at each node is considered in \cite{Liu13}. Steiner-tree packings are examined in \cite{Kannan13} for solving function computation jointly with multicasting, albeit without considering limitations on computation capacity of nodes. A line of work also deals with scaling laws of network computational capacity, cf.~\cite{Giridhar05} and followup papers.
	
	In dynamic setups, \cite{Liu13} examines the problem of function computation in cloud computing and use intuition from the Lagrangian relaxation to derive a dynamic queue-based algorithm. The work in \cite{Banerjee12} deals with the problem of computing a function of data generated at \emph{all} nodes in a network, a problem that is mainly motivated by sensor network applications. The authors relate the problem to the network broadcast (in the reverse manner) and they propose a scheme based on the \emph{Random Useful Policy} (adaptive broadcast policy \cite{Massoulie07}) to achieve maximum query rate. On the contrary, in our scenario where data stem (possibly) from a strict subset of the nodes, the corresponding (reverse) multicast method is not successful in general, indicating that the consideration of computation capabilities at all nodes is crucial for such a methodology.
	
	A common  underlying assumption, at least implicitly, in the aforementioned works is that network nodes can process data in an unrestricted way. Constraints on packet combinations are considered in the literature of \emph{processing networks} which is very much related to our work, see e.g. \cite{Zhao10} for a recent review. These networks model the industrial assembly of components, whereby the network blueprint determines where the combination of various types of components takes place. Recent works on allocation of resources and utility optimization in processing networks include the work in \cite{Jiang09}, where the use of ``dummy components'' is made to get around the processing restrictions, and \cite{Huang11}, where the authors advocate minimizing the drift of a suitably perturbed quadratic Lyapunov function. Our problem setup generalizes the processing networks framework in the following manner: instead of combining any two components of the same type (e.g. any bottle with any cork), here each query has a tag and we need to combine pairs of data with the exact same tags.
	
	\subsection{Our Contribution}
	
	We study the dynamic resource allocation problem that arises in network computation with the aim to achieve the maximum query response rate. We design a universal methodology and algorithm to solve this problem for a broad class of operations on data encountered in practice such as arithmetic, logical, database-related or other types of operations. We abstract the operation as ''summation'', with the understanding that it stands for any operation of that broad class.
	
	We consider a scenario with two data sources $1$ and $2$, and a stream of dynamically arriving queries, each of which seeks to compute the sum of a datum from source $1$ and a datum from source $2$ and deliver the sum to a destination. The process takes place in a communication network with diverse computation and bandwidth resources. The restriction to two sources and a query stream is deemed necessary for presentation purposes, but it will become apparent that the analysis in the paper can easily be extended to multiple sources and multiple queries via a multiclass queueing extension.
	
	We design algorithms that orchestrate utilization of computation and bandwidth resources by performing (\textit{i}) dynamic load balancing of computations on available nodes, (\textit{ii}) unprocessed (raw) and processed data routing from source nodes to computation nodes and from computation nodes to the destination respectively, (\textit{iii}) scheduling of data from different queries on communication links and computation nodes. The proofs of algorithm optimality require non-trivial modifications of the well-known Backpressure (BP) routing and scheduling, including computation thresholding for capturing the tag constraint, randomization for decoupling routing and computation, and the use of stochastic coupling.
	The contributions of our work are as follows.
	\begin{itemize}
		\item We formulate the problem of max-throughput distributed computation and derive necessary conditions for queue stability, which correspond to an upper bound on the maximum attainable query rate.
		\item For the optimal policy, we deploy our approach in stages. For a pre-specified computation node, we first derive the optimal policy under the restriction that the network infrastructure for communicating raw data is separated from the one for processed data. We then extend our approach to the unrestricted case. The optimal policy  involves \emph{Backpressure} for scheduling and routing and appropriate combining at computation nodes, and the optimality is derived through a novel queueing structure abstraction at those nodes.
		\item We extend to multiple computation nodes where computations need to be load balanced across the available options. The extended optimal policy is the one that is based on the \emph{join-the-shortest-sum-of-queues} rule.
	\end{itemize}
	
	The organization of the paper is as follows. In section \ref{sec:2}, we provide the model and assumptions. In section \ref{sec:Single_computation_node} we deploy our approach for a single computation node and we extend it to multiple possible computation nodes in \ref{sec:multiple}. Numerical results are presented in section \ref{sec:eval} and the paper is concluded in section \ref{sec:conclusion}.
	
	\section{Model and Problem Statement} \label{sec:2} 
	
	
	\subsection{Network, Resources and Query Streams}
	
	We consider a network abstracted as a graph $\mathcal{G}=(\mathcal{V}, \mathcal{E})$ where $\mathcal{V}$ is the set of nodes and $\mathcal{E}$ is the set of edges. We assume there exist two source nodes $s_1,s_2\in\mathcal{N}$ and a destination node $d$. 
	Edge $(m,l)\in\mathcal{E}$ between nodes $m$ and $l$ has a fixed capacity of $R_{ml}$ packets per slot. A network example is given in Fig. \ref{fig:network}.
	
	We study a \emph{stream} of queries, where each query concerns the computation of the sum of a datum from source $1$ and a datum from source $2$, while the network is agnostic to specificities of data.
	\footnote{An extension here is to consider networks that are aware of data specificities and can exploit them to improve performance; e.g.~use caching or multicast.} 
	This situation is abstracted as follows. Upon arrival of each query, a corresponding packet (datum) is generated at each of the two source nodes, and both packets are given the same \emph{tag}. These packets need to be summed somewhere in the network, and the result needs to be delivered to the destination $d$. Time is slotted, and at each slot $t$ there are $A(t)$ newly arrived queries belonging to the same stream, where the process $A(t)$ is assumed to be independent and identically distributed with time, with $\mean{A(t)}=\lambda$.
	
	Combination of packets corresponding to a query may take place in one among a subset of nodes, denoted by $\mathcal{N}_C=\{n_1,n_2,...,n_{N_C}\} \subseteq \mathcal{V}$; these are referred to as the \textit{computation nodes}. Node $n_i$ has computational capacity of $C_{n_i}$, measured in number of produced processed packets per slot, where each processed packet concerns the sum of two raw packets with the same tag when both are available to the computation node.
	
	\subsection{Operations and Embeddings}
	
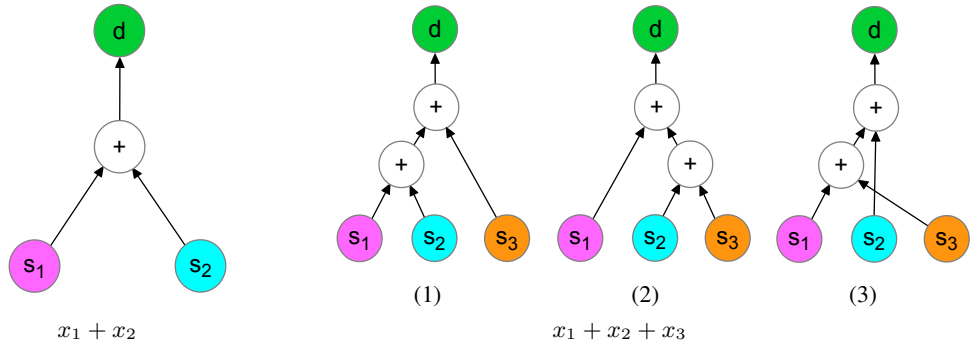
\begin{figure}[t!]
		\begin{center}
			\begin{overpic}[scale=0.5]{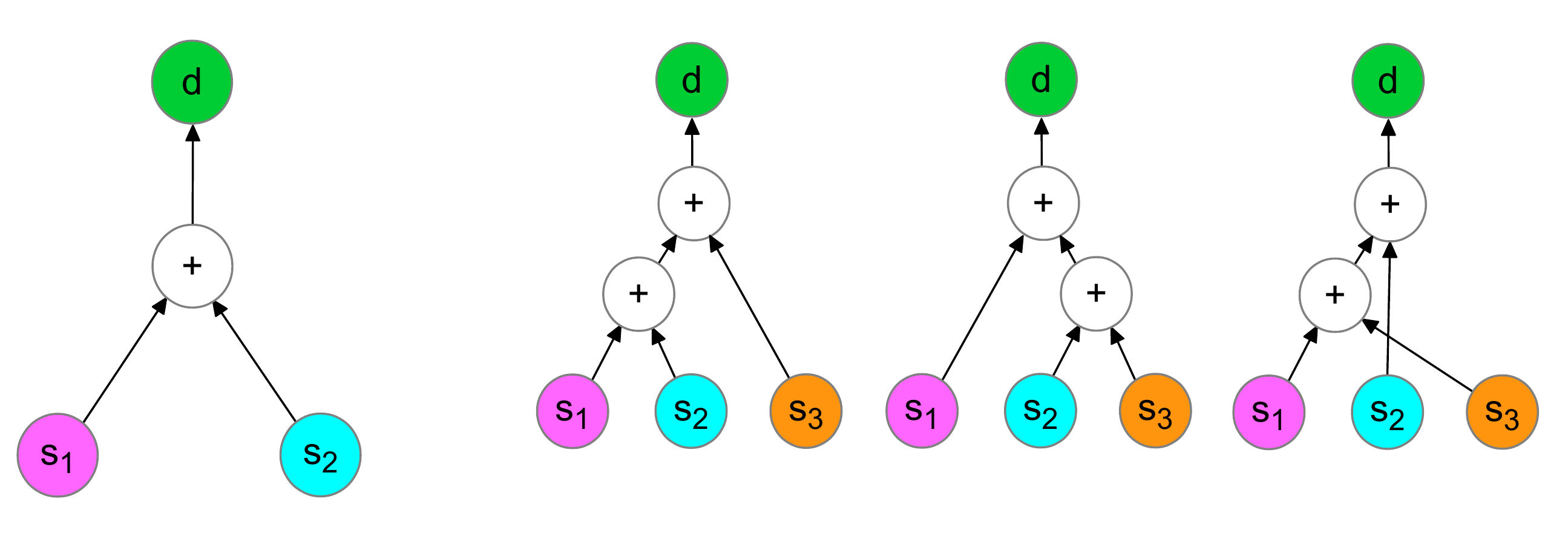}
				\put(6,-2){\footnotesize $x_1+x_2$}
				\put(56,-2){\footnotesize $x_1+x_2+x_3$}
				\put(42, 1.5){\footnotesize  (1)}
				\put(64, 1.5){\footnotesize  (2)}
				\put(86, 1.5){\footnotesize (3)}
			\end{overpic}
			\caption{Computation DAGs. In this paper we focus on a single operation with two sources and a unique computation DAG (shown left).} \vspace{-0.35in}
			\label{fig:dag}
		\end{center}
	\end{figure}

	For demonstration purposes, our analysis is focused on a simple operation $x_1+x_2$, but it is useful to discuss the generality of our model.
	Each computation task is associated with a set of sources 
	whose data are involved in the computation, and the operation to be performed on their data. For instance $x_1 + x_2 + x_3$ describes retrieval of one datum from each of the sources $1,2,3$ and their addition. From the network computation point of view however, the description of the operation is completely specified only when we are given the entire order of how data are combined. One way to provide such a description is the so-called \textit{computation graph}, which is a directed acyclic graph (DAG) whose nodes are the sources, the destination, and the operations themselves. The ordering of nodes in this graph gives a description of the operation. Some operations are associated with a unique computation DAG while some others do not. For example, the operation $x_1 + x_2$ is associated with a unique DAG with nodes $1$, $2$ and ``$+$'' denoting the summation, see Figure~\ref{fig:dag}-(left). On the other hand, operation $x_1 + x_2 + x_3$ has more than one computation DAGs, each of which stems from the outcome of the associativity property of the addition operator. In this work, we will assume that each task is associated with a \emph{unique} computation DAG.
	
	An embedding of the computation graph on the network graph is a mapping of DAG operation nodes to computation nodes of the network. Prior work has studied the problems of finding an embedding that minimizes delay and cost and has shown that they are both NP-complete problems \cite{Vyavahare14}. In this paper we focus on one query stream that can be computed at $N_C$ nodes, and hence there are $|\mathcal{N}_C|$ possible embeddings of our computation DAG onto $G$. Instead of finding the best embedding, we use all embeddings available to load balance computation and achieve the maximum attainable query rate.
	
	We remark that our analysis can be generalized to multiple query streams and multiple computation DAGs using a multiclass queueing approach that we omit here for brevity. Moreover, the case we are studying entails all the complexity that arises from integrating routing and computation.
	
	\subsection{Queueing Model}\label{ssec:Queueing_Model}
	
	Data packets at each node may be (\textit{i}) unprocessed (raw) source data on their way from the source to the computation node, or (\textit{ii}) processed data on their way from the computation node to the destination. We introduce a packet classification with respect to the computation node that a particular packet will be (or was) computed. The raw packets can be further classified according to the source where they stem from. To capture all packet classes we define the following queues:
	\begin{itemize}
		\item $\mathcal{Q}_k^{(i,n)}(t), i=1,2$: Data queue at node $k$ containing raw packets generated at node $s_i$ that have to be computed at node $n$; ${Q}_k^{(i,n)}(t)$ denotes its length. We make the convention that ${Q}_n^{(i,n)}(t)=0$.
		\item $\mathcal{X}_n^{(i)}(t), i=1,2$: Computation queue at node $n$ containing raw packets generated at node $s_i$ that have to be computed at \emph{this node}; ${X}_n^{(i)}(t)$ denotes its length.
		\item $\mathcal{Q}_k^{(0,n)}(t), i=1,2$: Data queue at $k$ containing processed packets from computing node $n$, that have to be delivered at the destination node; ${Q}_k^{(0,n)}(t)$ denotes its length. We make the convention that $Q^{(0,n)}_d(t)=0, \forall n\in\mathcal{N}_C$.
	\end{itemize}
	
	Moving packets between queues corresponds to control decisions to be taken each slot:
	\begin{itemize}
		\item The set of raw packets with tags $\mathcal{U}_{mk}^{(i,n)}(t)$ originated from node $s_i$, destined to computation node $n$, that are transmitted from node $m$ to node $k$; ${U}_{mk}^{(i,n)}(t)$ is the number of packets of this decision. We allow to allocate more service than packets waiting in the queue, in which case "zero" packets are transmitted (these packets will be dropped at the other side of the link).
		\item The pairs of raw packets to be combined at each computation node $n$. Let $\mathcal{Z}_n(t)$ be the set of corresponding tags and $Z_n(t)$ be the number of combined packets.
		\item The set of processed packets, combined at node $n$, $\mathcal{U}_{mk}^{(0,n)}(t)$ to be transmitted from node $m$ to node $k$; ${U}_{mk}^{(0,n)}(t)$ is the number of such packets.
	\end{itemize}
	
	We have the following constraints. The total number of transmitted packets over a link are limited by link capacity 
	\begin{equation}
		\sum_{\substack{i\in\{0,1,2\},\\n\in\mathcal{N}_C}}\left(U_{ml}^{(i,n)}(t) + U_{lm}^{(i,n)}(t)\right) \leq R_{ml},~~\forall (ml)\in\mathcal{E}.\label{eq:capcon}\vspace{-0.08in}\end{equation}
	Further, the number of combined pairs cannot exceed the computation capacity or any of the individual raw packet queue lengths,
	\begin{equation}
		0\leq Z_n(t)\leq \min\left[C_n, X_n^{(1, n)}(t), X_n^{(2,n)}(t)\right], ~~\forall n\in\mathcal{N}_C\,.\label{eq:comcon1}
	\end{equation}
	Moreover, a pair of packets can be combined only if both packets with the same tag have already arrived at the computation node, i.e.,
	\begin{equation}
		\mathcal{Z}_n(t) \subseteq \mathcal{X}_n^{(1)}(t)\cap \mathcal{X}_n^{(2)}(t), ~~\forall n\in\mathcal{N}_C.\label{eq:comcon}
	\end{equation}
	We point out that the last constraint involves consideration of packet tags and would complicate the description of the system state. However, our approach will be to define a simpler system state with queue lengths only, and then establish that the considered policies indeed satisfy \eqref{eq:comcon}.
	
	We define the set of permissible policies in our system $\Pi_C$ as mappings of the network state (queue lengths) to control variables for routing $U_{ml}^{(i,n)}(t)$ and computation $Z_n(t)$, subject to capacity  and computation constraints \eqref{eq:capcon}-\eqref{eq:comcon}.
	
	\subsection{Problem Formulation}
	
	We say that \emph{the system is stable} under a policy $
	\pi$ if all queues in the system are strongly stable, i.e. if
	\begin{align}\nonumber
		\limsup\limits_{T\rightarrow\infty}\frac{1}{T}\sum\limits_{t=1}^{T}\EE\left\{Q_k^{(i,n)}(t)\right\}<\infty, ~\forall i\in\{0,1,2\}, \forall k\in\mathcal{N} \\ \nonumber
		\limsup\limits_{T\rightarrow\infty}\frac{1}{T}\sum\limits_{t=1}^{T}\EE\left\{X_n^{(i)}(t)\right\}<\infty,~~ \forall i\in\{1,2\},~ \forall n\in\mathcal{N}_C.\nonumber	
	\end{align}
	We are interested to find the  maximum attainable query rate, $\lambda^*$ that can be delivered by some policy in the class $\Pi_C$ subject to system stability, as well as to find a policy $\pi^*\in\Pi_C$ that stabilizes the system for \emph{every} query rate $\lambda<\lambda^*$. 
	
	It is important to note that strong (or at least steady-state) stability of all queues is actually necessary to ensure that all computations are made and results are delivered to the destination. Indeed, if some queues are only mean-rate- or rate-stable (these are weaker notions of stability), there may be a growing number of queries in time that are never executed.
	
	
	\section{Single Computation Node}\label{sec:Single_computation_node}
	
	We begin our analysis by fixing attention to one computation node, say node $n$. This special case contains the crux of the problem, which is to deal with (\textit{i}) the challenging constraint \eqref{eq:comcon}, and (\textit{ii}) integration routing and computation. 
	
	\subsection{Query Rate Upper Bound $\lambda^*$}\label{ssec:single_upper_bound}
	
	First we revisit the standard multicommodity flow problem. For a set of commodities $\mathcal{C}$, consider the multicommodity flow feasibility region $\Lambda_{\mathcal{G}}(\mathcal{C})$ of network $G$ which is defined as the set of arrival rate vectors $(\lambda^{(c)})_{c\in\mathcal{C}}$ for which \emph{there exists a feasible flow that successfully decomposes the arrivals}. Feasibility in this case includes, (\textit{i}) flow conservation constraints $\forall c\in\mathcal{C}$
	\begin{equation}\label{eq:flow_consv}
		\sum_{k\in \text{OUT}(m)} f_{mk}^{(c)}- \sum_{k\in \text{IN}(m)} f_{km}^{(c)}=\left\{\begin{array}{lll}
			\lambda^{(c)} & m=\text{src.} \\
			-\lambda^{(c)} & m=\text{dest.} \\
			0 & \text{otherwise,}
		\end{array}\right.
	\end{equation}
	(\textit{ii}) capacity constraints,
	\begin{equation}\label{eq:cap_const}
		\sum_{c\in\mathcal{C}} f_{mk}^{(c)}\leq R_{mk},~~\forall (m,k) \in \mathcal{E},
	\end{equation}
	and (\textit{iii}) standard flow constraints,
	\begin{equation}\label{eq:flow_const}\left\{
		\begin{array}{ll}
			f_{mk}^{(c)}=0 & m=\text{dest.} \\
			f_{mk}^{(c)}\geq 0 &  \text{otherwise}
		\end{array}\right.~~\forall (m,k) \in \mathcal{E},~\forall c\in\mathcal{C}.
	\end{equation}
	The feasibility region, given by the convex polytope
	\[
	\Lambda_{\mathcal{G}}(\mathcal{C})=\{(\lambda^{(c)})_{c\in\mathcal{C}}~|~\eqref{eq:flow_consv}-\eqref{eq:flow_const}\}
	\]
	is also the set of arrival rates for which the system with dynamic routing policies (for the same network and commodity setting) is stable, under mild assumptions for arrival processes \cite{Georgiadis06}.
	
	Consider now the standard multicommodity routing problem with three commodities $\mathcal{C}_3=\{(s_1,n),(s_2,n),(n,d)\}$, and corresponding feasibility region $\Lambda_{\mathcal{G}}( \mathcal{C}_3)$. We have:
	
	\begin{theorem}\label{th:upperbound}
		For a query stream  from sources $s_1,s_2$, destined to $d$ and computed at node $n$, the following are necessary conditions for system stability:
		\[
		(\lambda,\lambda,\lambda)\in \Lambda_{\mathcal{G}}( \mathcal{C}_3),\quad \lambda\leq C_n.
		\]
	\end{theorem}
	\begin{IEEEproof}
		Recall that the policy set $\Pi_C$, whose stability region we are interested in bounding, involves the challenging constraint \eqref{eq:comcon}. To obtain the upper bound on the performance of this class, we relax this constraint in the following way. Consider a new set of control policies $\Pi$, whereby we select routing variables $U_{ml}^{(i,n)}(t)$ subject to instantaneous capacity constraint $R_{ml}$, and computation variable $Z_n(t)$ subject to capacity $C_n$, while constraint \eqref{eq:comcon} is relaxed and any two raw packets can be combined. Note that policies in set $\Pi\setminus \Pi_C$ may interleave raw data from different queries, which harms the system.
		
		The conditions in the statement of the theorem are necessary for stability for any policy $\pi\in \Pi$; the first condition is necessary for routing all raw packets from the sources to $n$  and the combined packets from $n$ to the destination and the second is necessary for performing all computations. Then the proof is complete by noting that if $\pi\in\Pi_C$ then $\pi\in\Pi$ as well--this is due to the fact that $\Pi$ is obtained from $\Pi_C$ by relaxing a constraint, hence it also includes policies that satisfy the constraint.
	\end{IEEEproof}
	
	Equivalently, for a stable system the query rate is upper bounded by
	\[
	\lambda^*=\max_{\substack{(\lambda,\lambda,\lambda)\in \Lambda_{\mathcal{G}}( \mathcal{C}_3)\\ \lambda<C_n.}}\lambda.
	\]
	In the next subsections we propose policies in $\Pi_C$ that provably stabilize any $\lambda<\lambda^*$, thus establishing that in fact $\lambda^*$ is the maximum query rate.

	\subsection{Achieving maximum sustainable query rate in Non-overlapping Networks}\label{ssec:single_disjoint}
	
\begin{figure}[t!]
		\begin{center}
			\hspace{-0.8in}\begin{overpic}[scale=0.75]{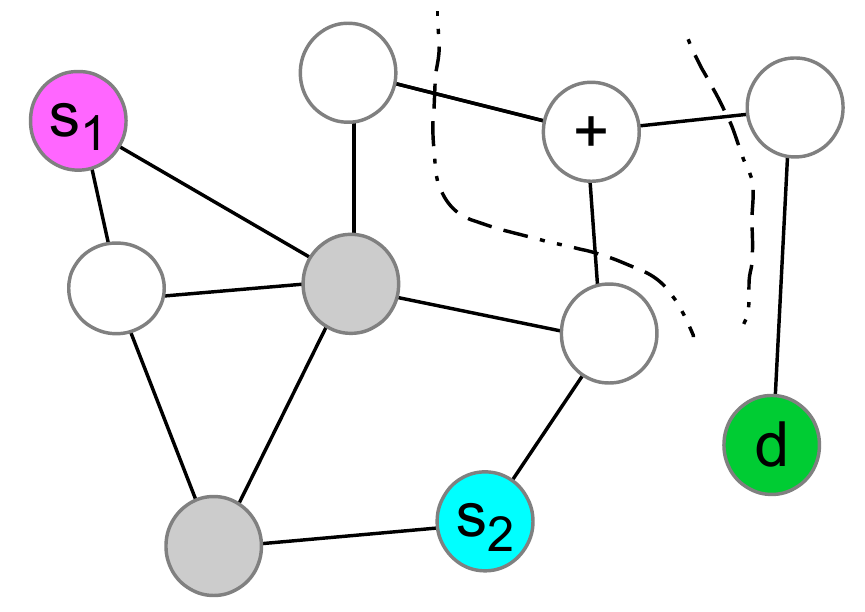}
				\put(1,-3){\small \textbf{network with raw data}}
				\put(93,43){\small \textbf{network with}}
				\put(93,36){\small \textbf{processed data}}
			\end{overpic}
			\caption{Example where the networks of raw and combined data are non-overlapping.
			} \vspace{-0.3in}
			\label{fig:nonoverlap}
		\end{center}
	\end{figure}

	To ease exposition, we examine first the special case where the network connecting the sources to the computation node is non-overlapping with the network connecting the computation node to the destination, see Fig.~\ref{fig:nonoverlap}.
	When this is true, the networks containing $s_1, s_2, n$ and $n,d$ do not have common nodes (except $n$) and hence the routing is decoupled from computation, and the only remaining complication is the challenging constraint \eqref{eq:comcon}. 
	
	We first consider the case where the destination is also the node that performs the computations. The results will be generalized in a straightforward way at the end of the section. Consider the following policy $\pi_1$:
	\begin{itemize}
		\item The controls $U_{mk}^{(i,n)}(t)$ are obtained by applying \emph{BP routing and scheduling}; see the text box and \cite{Tassulas92}, \cite{Neely05} for details.
		\item At node $n$, let $P_n(t)$ denote the number of packet tags that satisfy \eqref{eq:comcon}. Combine the maximum number of available paired packets (i.e. $\min\{P_n(t),C_n\}$ pairs).
	\end{itemize}
	
	\noindent\fbox{
		\parbox{18.0cm}{
			{\center \textbf{Review on Backpressure (BP) Routing and Scheduling \cite{Tassulas92}:}}
			BP is a distributed dynamic algorithm that reacts on current data queue backlogs and decides on the number of packets to be routed across each link in order to balance queues. For wired multicommodity routing, the BP algorithm to be executed at each link/slot is:
			
			\noindent Select the commodities that maximize differential backlog:
			\[c_{mk}^*(t)=\argmax_{c\in\mathcal{C}}\left|Q_m^{(c)}(t)-Q_k^{(c)}(t)\right|,~\forall (m,k)\in\mathcal{E}\]
			\noindent Route any $U_{ij}^{(c)}(t)$ packets from commodity $c$, where:
			\[
			U_{mk}^{(c)}(t)=\left\{\begin{array}{ll}
			R_{mk} & \text{if}~ c=c_{mk}^*(t)~\text{and}~ Q_m^{(c)}(t)>Q_k^{(c)}(t) \\
			0 & \text{otherwise.}
			\end{array}\right.
			\]
			BP is a maximum-throughput policy for wired multicommodity routing.
		}
	}\\
	
	The main result of this section follows:
	
	\begin{theorem}\label{th:optimality_disjoint}
		For non-overlapping networks and a query stream computed at any one node $n$, policy $\pi_1$ is stable for any query rate $\lambda<\lambda^*$.
	\end{theorem}
	We prove the result by tackling constraint \eqref{eq:comcon}.
	To avoid tracking packet tags, define $X(t)$ as the number of raw packets in the system excluding node $n$, and $S_n^{(1)},S_n^{(2)}$ as the number of raw packets with unique tags at node $n$, whose counterpart is still in $X(t)$. Recall that $P_n(t)$ is the number of raw pairs with same tag waiting at node $n$. Breaking down the packets at node $n$ we have
	\begin{align*}\label{eq:condition_Cn_ex}
		X_n^{(1)}(t)+X_n^{(2)}(t)&=2P_n(t)+S_n^{(1)}(t)+S_n^{(2)}(t)\notag\\
		&\leq 	2P_n(t) + X(t)
	\end{align*}
	where inequality becomes strict equality if all packets $X(t)$  have different tags.
	Thus the number of pairs waiting at $n$ to be computed is lower-bounded by
	\[
	P_n(t)\geq \frac{X_n^{(1)}(t)+X_n^{(2)}(t)-X(t)}2,
	\]
	from which we observe that, if
	\begin{equation}\label{eq:condition_Cn}
		X_n^{(1)}(t)+X_n^{(2)}(t)\geq 2C_n + X(t)\,,
	\end{equation}
	then there are at least $C_n$ pairs of packets with the same tag in the computation node. 	Hence we may directly replace constraint \eqref{eq:comcon} with the restriction \eqref{eq:condition_Cn}. The idea behind the proof is to use a slightly different restriction from \eqref{eq:condition_Cn}, where, instead of using $X(t)$ (which is correlated with network events), we make use of a large constant $\overline X$.
	
	Consider a policy $\pi_1'$ that works as follows:
	\begin{itemize}
		\item The controls $U_{mn}^{(i,n)}(t)$ are obtained by applying backpressure routing and scheduling, as in $\pi_1$.
		\item At node $n$: choose the computations as follows:
		\[
		Z_n(t)=\left\{\begin{array}{ll}
		C_n & \text{if} ~~ X_n^{(1)}(t)+X_n^{(2)}(t)\geq 2C_n + \overline{X}\\
		0 & \text{otherwise}
		\end{array}\right.
		\]
	\end{itemize}
	The main idea is to prove the statement of Theorem \ref{th:optimality_disjoint} for policy $\pi_1'$ and then prove the theorem itself by showing that $\pi_1'$ is stochastically dominated by $\pi_1$. First, we prove that $\pi_1'$ satisfies the largest possible query rate:
	\begin{lemma}\label{le:threshold_based_disjoint}
		For any $\lambda<\lambda^*$, there exists a threshold $\overline{X}$ that depends on the  parameters $\mathcal{G},C_n,\lambda$, such that the network is stable under policy $\pi_1'$.
	\end{lemma}
	
	\begin{IEEEproof}
		We give here a sketch of the proof. The complete version of this proof as well as all other proofs of results here is in the Appendix. Since BP routing stabilizes the network queues, and leads to a stationary distribution for their lengths, we pick a large enough value $\overline X$ so that the probability that $X(t)>\overline X$ is made sufficiently small. Then we consider the $T$-slot drift for the computation queues and show that they are also stable, using the fact that the restriction \eqref{eq:condition_Cn} is violated only very rarely.	
	\end{IEEEproof}
	
	Policy $\pi_1'$ achieves the computational capacity of the network but has some shortcomings. First, computation of an appropriate threshold $\overline{X}$ can be tedious and requires the statistics of the query arrival processes. Second, policy $\pi_1'$ adds delays in delivering the result to destination, since due to large $\overline X$ used, the computing node does not perform any computations until many packets have arrived. Both issues above are resolved by policy $\pi_1$, which does at least as good as $\pi_1'$ in terms of stability, as implied by the following result:
	
	\begin{lemma}\label{le:dominance_disjoint}
		For all thresholds $\overline{X}$, we have
		\[X_n^{(i), \pi_1}(t) \leq_{\text{st}}X_n^{(i), \pi_1'}(t),~~ \forall i\in\{1,2\}.\]
	\end{lemma}
	\begin{IEEEproof}
		We compare what happens when both policies start with the same network state at time $t=0$ and the sample paths of routing decisions and arrival processes are the same. The sample paths of the computations made, however,  are different between the two policies. Let $t'$ the time slot index where the constraint for $\pi_1'$ to be active holds for the first time. The main point is that, in the meantime, policy $\pi_1$ can serve packets, therefore all packets that will get combined by $\pi_1'$ at $t'$ will have been combined earlier, ot at $t'$ if its the first time pairs of packets appear. That is, the $C_n$ packets that are combined by $\pi_1'$ have either been combined already by $\pi_1$ if some of these were paired earlier than $t'$, or they have been combined by $\pi_1$ if all of these are paired on $t'$, so $X_n^{(i), \pi_1}(t')\leq X_n^{(i), \pi_1'}(t')$. In addition, since for $t<t'$ no packets are served by $\pi_1'$, while maybe served by $\pi_1$, so $X_n^{(i), \pi_1}(t)\leq X_n^{(i), \pi_1'}(t), \forall t<t'$. We can show, using similar arguments that this inequality holds also for $t>t'$.
		
		Since this is proven for every sample path for arrivals and routing decisions, the statement in the Lemma follows.   		
	\end{IEEEproof}
	
	Finally, the proof of Theorem \ref{th:optimality_disjoint} follows from the above lemmas.
	
	\subsection{Achieving maximum sustainable query rate in Overlapping Networks}\label{ssec:single_overlapping}
	
	Next, we relax the topological restriction and we allow processed packets to compete with raw packets for the same network resources. The arising technical problem here is the dependence of network state on the outputs of the computation node, which means that certain steps used to prove Lemma~\ref{le:threshold_based_disjoint} can not be used directly.
	
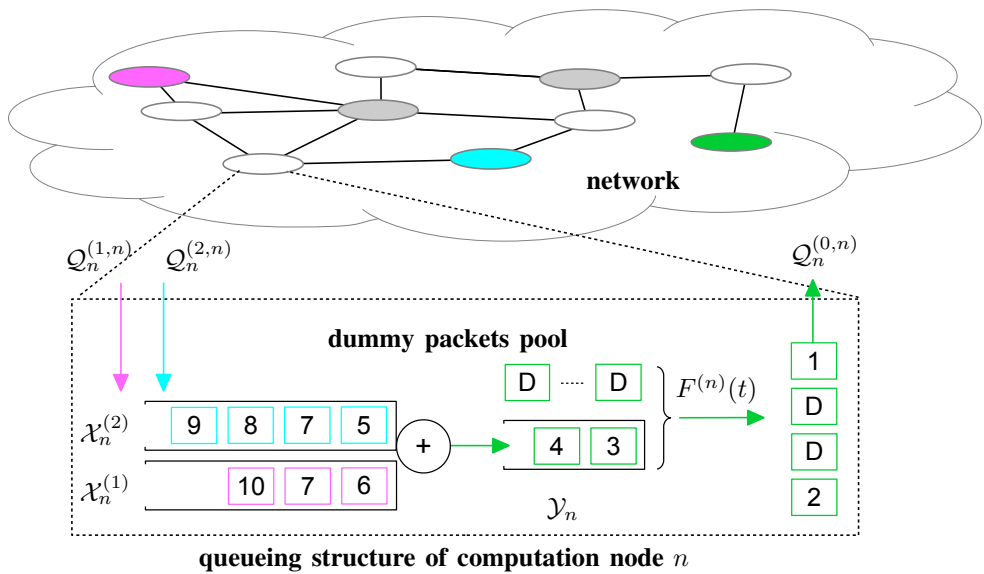
\begin{figure}[h!]
		\begin{center}
			\begin{overpic}[scale=0.6]{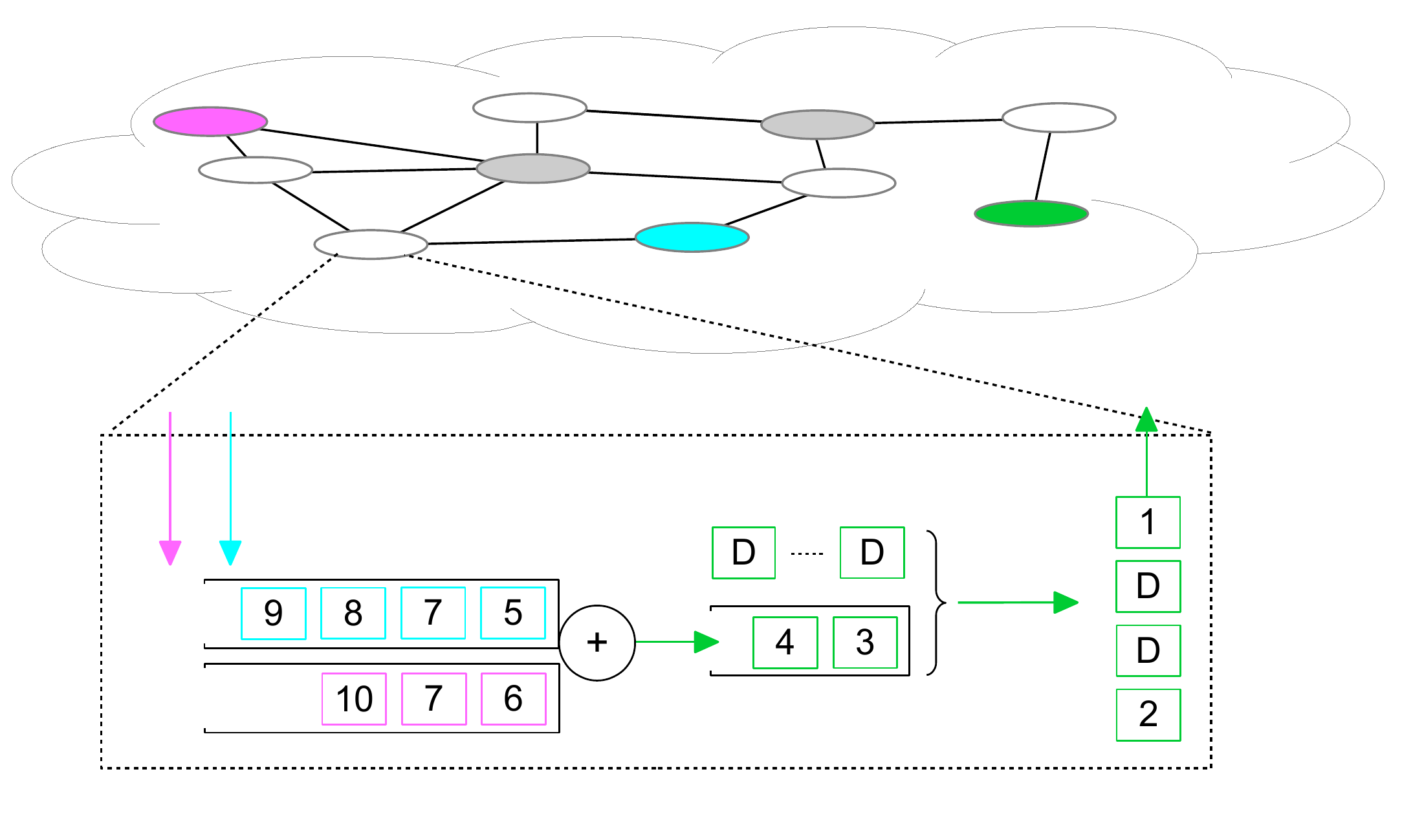}
				\put(33,24){\small \textbf{dummy packets pool}}
				\put(59,40){\small \textbf{network}}
				\put(20,2){\small \textbf{queueing structure of computation node} $n$}
				\put(68,19){\small $F^{(n)}(t)$}
				\put(55,7){\small $\mathcal{Y}_n$}
				\put(8,8.5){\small $\mathcal{X}_n^{(1)}$}
				\put(8,14.5){\small $\mathcal{X}_n^{(2)}$}
				\put(79.5,32.5){\small $\mathcal{Q}_n^{(0,n)}$}
				\put(6.5,32){\small $\mathcal{Q}_n^{(1,n)}$}
				\put(16.5,32){\small $\mathcal{Q}_n^{(2,n)}$}
			\end{overpic}
			\caption{Illustration of queueing structure for computation  node $n$. Numbered packets are either raw (red and blue) or processed (purple) useful packets, while packets noted with ``D'' are dummy packets. At slot $t$, $F_n(t)$ packets depart the queue $\mathcal{Y}_n$ and arrive at the network queue $\mathcal{Q}_n^{(0,n)}$, where $F_n(t)$ potentially includes both dummy and useful packets.}
			\vspace{-0.2in}
			\label{fig:computing_node}
		\end{center}
	\end{figure}

	In order to overcome this technicality, we randomize the output of the computation node by purposefully inserting dummy packets at random. In particular, we introduce an additional queue $\mathcal{Y}_n(t)$ in the computing node that keeps the results of the computation and forwards $F_n(t)$ processed packets  to the queue $\mathcal{Q}_n^{(0,n)}(t)$, where $F^{(n)}(t)$ includes both useful and dummy packets. The network then treats these dummy packets as real ones, delivering them to the destination. This structure is shown in Fig.~\ref{fig:computing_node}.  
	
	At this point we refer to the old concept of regulators, first proposed in \cite{Humes94} to fix the arrivals of an intermediate queue to be equal to the source. Later regulator queues were used for wireless scheduling in \cite{Wu07,Joo09}, although in these approaches the knowledge of the arrival rate is required.
Our technique is slightly different since it does not require this knowledge.
	
	The proposed structure above leads to a situation where the number of packets going into $\mathcal{Q}_n^{(0,n)}(t)$ does not depend on the network queues, which greatly helps with the analysis. To perform the analysis, however, we additionally need to design carefully the number of generated dummy packets to ensure that the adjusted packet rate remains inside region $\Lambda_{\mathcal{G}}(\mathcal{C}_3)$, and that the virtual queue $\mathcal{Y}(t)$ is served enough to be strongly stable. These two properties then imply that the useful processed packets will eventually be delivered.
	
	We will consider the policy $\pi_2$. At each time slot $t$:
	\begin{itemize}
		\item Controls $U_{mk}^{(i,n)}(t)$ are obtained by applying backpressure routing and scheduling.
		\item At node $n$, all paired packets that can be combined (i.e. at most $C_n$ pairs) are combined and pushed to the queue $\mathcal{Y}_n(t)$.
		\item $F^{(n)}(t)$ packets are pushed from the queue $\mathcal{Y}_n(t)$ to the queue $\mathcal{Q}_n^{(0,n)}(t)$. $F^{(n)}(t)$ is a random variable with mean $\lambda'\in(\lambda, \lambda^*)$. If there are not enough packets, then dummy packets are used. These dummy packets are routed exactly as the normal processed packets.
	\end{itemize}
	For the process that serves queue  $\mathcal{Y}_n(t)$  
	we use 
	\begin{equation}\label{eq:service_Y_single}
		F^{(n)}(t)=(1+B^{(n)}(t))A(t)
		,\end{equation}
	where $B^{(n)}(t)$ is a Bernoulli random variable, independent of everything in the network, with success probability $\epsilon_B$; this parameter can take an arbitrarily small positive value.  	
	
	The main result of this subsection is that the policy described above achieves a computation rate arbitrarily close to the upper bound:
	
	\begin{theorem}\label{th:optimality_general}
		For one query stream computed at $n$, policy $\pi_2$ is stable for any query rate $\lambda<\left(1-\frac{\epsilon_B}{1+\epsilon_B}\right)\lambda^*$.
	\end{theorem}
	
	This theorem can be proved in a similar way as Theorem~\ref{th:optimality_disjoint}, by defining a policy $\pi_2'$ that is the same as $\pi_2$, except that it does computations only if $X_n^{(1)}(t)+X_n^{(2)}(t)\geq 2C_n + \overline{X}$. Similarly to the previous subsection, we first prove stability under $\pi_2'$:
	
	\begin{lemma}\label{le:threshold_based_general}
		For any $\lambda<\left(1-\frac{\epsilon_B}{1+\epsilon_B}\right)\lambda^*$, there is a threshold $\overline{X}$ such that the the network is stable under policy $\pi_2'$.
	\end{lemma}
	
	\begin{IEEEproof}[Proof Outline]
		The crucial observation is that, due to the randomized input to queue $\mathcal{Q}_n^{(0,n)}(t)$ and the use of dummy packets,  $\left({Q}_k^{(0,n)}(t)\right)_{k\in\mathcal{N}}$ do not depend on the state of the queues with raw packets.For the network excluding node $n$, then, it is as if we had three commodities: $s_1\rightarrow n, s_2\rightarrow n, n\rightarrow d$ with rates $(\lambda,\lambda,(1+\epsilon_B)\lambda$ inside the stability region of the system and correlated arrival processes, but i.i.d. in time. This implies that $Q_k^{(i,n)}(t), \forall i\in{0,1,2},\forall k\in\mathcal{V}\setminus \{n\}$  are strongly stable under $\pi_2'$ (since backpressure routing is applied).
		
		In addition, strong stability implies steady-state stability of the aforementioned queues, so $X(t)$ has a steady-state distribution with zero limit as it goes to infinity. We can then apply the same methodology as in the proof of Theorem \ref{th:optimality_disjoint} to show that there exists a threshold $\overline{X}$ for which $\mathcal{X}_n^{(i)}(t), i \in\{1,2\}$ are strongly stable.
		
		The final step is to show that queue $Y_n(t)$ is stable as well. Indeed, since  $\mathcal{Q}_n^{(i)}(t), i \in\{1,2\}$ are stable, the input to this queue is a Markov modulated process with mean $\lambda$. Since the service process of $\mathcal{Y}_n(t)$ is i.i.d. over time with mean $\left(1-\frac{\epsilon_B}{1+\epsilon_B}\right)\lambda^*>\lambda$, strong stability of $\mathcal{Y}_n(t)$ follows.
	\end{IEEEproof}
	
	Then, we prove that the threshold is not really needed, in the same way as Lemma \ref{le:dominance_disjoint}.
	\begin{lemma}\label{le:dominance_general}
		For all thresholds $\overline{X}$, $X_n^{(i), \pi_2}(t) \leq_{\text{st}}X_n^{(i), \pi_2'}(t), \forall i\in\{1,2\}$
	\end{lemma}
	
	Theorem \ref{th:optimality_general} is then a consequence of Lemmas \ref{le:threshold_based_general} and \ref{le:dominance_general} and the observation that the number of packets (including dummy ones) injected in queue $\mathcal{Q}_n^{(0,n)}(t)$ at every time slot $t$  is the same for both policies $\pi_2, \pi_2'$. As a final remark, note that policy $\pi_2$ can achieve any fraction of the computational capacity upper bound of Theorem \ref{th:upperbound} by requiring only access to the number of requests at every slot. In fact it can be proven that even delayed information about the number of requests is sufficient for this policy to achieve any fraction of the computational capacity of the network \cite[\S 4.7]{Georgiadis06}.
	
	In summary, we have shown that for one stream with two sources and one destination, where the streams need to be combined at specified node $n$, the maximum query rate is characterized by  Theorem~\ref{th:upperbound}, and achieved in a distributed way by policy $\pi_2$. In the next section we generalize to multiple computation nodes.
	
	
	\section{Multiple Possible Computation Nodes} \label{sec:multiple}
	
	In this section, we consider the more general model where the summation can take place at any one of the $\mathcal{N}_C=\{n_1,\dots,n_{N_C}\}$ computation nodes.
	
	\subsection{Query Rate Upper Bound $\lambda^*$}
	
	In this scenario, we have three commodities for every one of the $N_C$ computation nodes, hence we need to define the set $\widetilde{\mathcal{C}}_{3}$ with $3N_C$ unicast commodities, as follows: there are three commodities for each computation node $n \in \mathcal{N}_C$, $(1,n)$ delivering packets from $s_1$ to $n$, $(2,n)$ delivering packets from $s_2$ to $n$, and $(n,d)$ delivering packets from $n$ to $d$. Consider the multicommodity flow with rates $\bm{\lambda}=(\lambda_1^1,\lambda_1^2,\lambda_1^0,\dots,\lambda_{N_C}^1,
	\lambda_{N_C}^2,\lambda_{N_C}^0)$, where $\lambda_m^1=\lambda_m^2=\lambda_m^0=\lambda_m$, and
	\[
	\sum_{m\in \mathcal{N}_C}\frac{\lambda_{m}}{\lambda}=1,
	\]
	i.e., the quantities $(\frac{\lambda_{m}}{\lambda})_{m\in\mathcal{N}_C}$ can be seen as the time-share coefficient for queries computed at node $m$. Then we have the following upper bound for the query rate:
	\begin{theorem}\label{th:upperbound2}
		For a query stream with sources $s_1,s_2$, destined to $d$ and computed at the set of computation nodes $\mathcal{N}_C$, the following is a necessary condition for stability:
		\[
		\bm{\lambda}\in \Lambda_{\mathcal{G}}( \widetilde{\mathcal{C}}_{3}),\quad \lambda_{m}\leq C_m,~\forall m\in\mathcal{N}_C, ~~ \sum_{m\in \mathcal{N}_C}\lambda_{m}=\lambda.
		\]
	\end{theorem}
	
	The upper bound characterized by Theorem~\ref{th:upperbound2} can be actually achieved arbitrarily close by a dynamic policy, as discussed in the next subsection.
	
	\subsection{Achieving maximum sustainable query rate  with Multiple Computation Nodes}
	
	In addition to the queues specified in \ref{ssec:Queueing_Model}, we need to define $N_C$ other queues, denoted with $H_n(t)$, whose role is to ensure that each computing node does not receive more computational load than its capacity. Queues $H_n(t)$ evolve as
	\[
	H_n(t+1) = \left[H_n(t) + \tilde{A}^{(n)}(t) - C_n\right]^+\,.
	\]
	
	The dynamic policy $\pi_3$ we consider here is the following:
	\begin{itemize}
		\item \textbf{Load Balancing:} At each slot, choose $n^*(t)$ equal to
		\begin{equation} \argmin\limits_{n\in\mathcal{N}_C}\left[(1+\epsilon_B)Q_n^{(0,n)}(t)+\sum_{i=1,2}Q_i^{(i,n)}(t) + H_n(t)\right]
		\end{equation}
		where $\epsilon_B\in(0,1)$ is a control parameter. Then the newly arrived queries are assigned to the class that corresponds to this computation node,
		\begin{equation}
			\tilde{A}^{(n)}(t)=\begin{cases}
				A(t), n=n^*(t)\\
				0, \text{otherwise.}
			\end{cases}
		\end{equation}
		\item \textbf{Routing and scheduling:} Use BP over class pairs. For every link $(m,k)\in\mathcal{E}$ choose the class pair
		\begin{align}\nonumber
			&\left(i_{mk}^*(t), n_{mk}^*(t)\right)=\argmax\limits_{\substack{i\in\{0,1,2\}\\ n\in\mathcal{N}_C}}\left|Q_{m}^{(i,n)}(t)-Q_k^{(i,n)}(t)\right|,	
		\end{align}
		where $i_{mk}^*(t)$ is the best class of packets between raw packets and processed packets, and $n_{mk}^*(t)$ is the best class of packets w.r.t. the computation node. Then choose the routing variables as,
		\begin{align*}
			U_{mk}^{(i,n)}(t)=\left\{\begin{array}{ll}
				R_{mk} & \text{if }~~(i,n)=(i^*(t),n^*(t))\\
				0 & \text{otherwise.}	\end{array}\right.
		\end{align*}
		\item \textbf{Computation:} At every node $n\in\mathcal{N}_C$, all possible computations are done. If there are more pairs than the computation capacity of this node, then $C_n$ pairs are selected using any tie breaking rule (e.g. priority can be given to the oldest queries).
		\item \textbf{Randomization with dummy packets:} $F^{(n)}(t) = \tilde{A}^{(n)}(t)\left(1 + B^{(n)}(t)\right)$ packets resulting from a computation are pushed to queue $\mathcal{Q}_n^{(0)}(t)$, where $B^{(n)}(t)$ are an i.i.d. Bernoulli random variables with mean $\epsilon_B$. If there are not enough processed packets available at queue $\mathcal{Y}_n$, dummy packets are used.
	\end{itemize}
	
	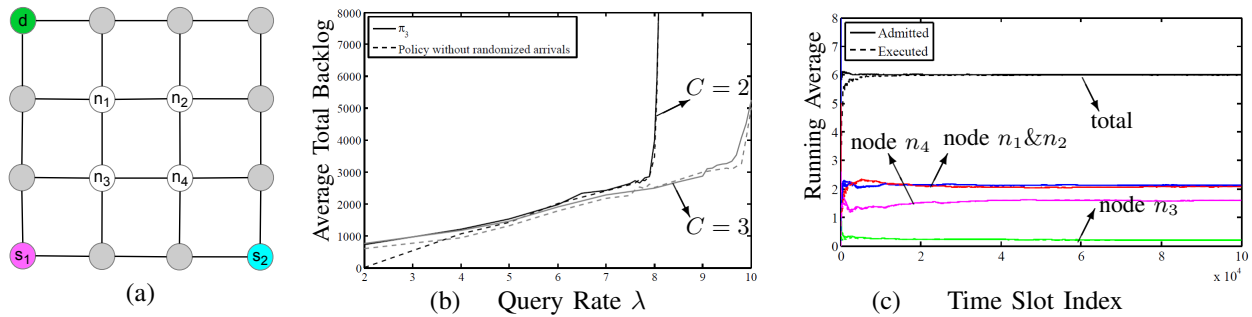
\begin{figure*}[t!]
		\centering
		\hspace{0.03in}
		\begin{overpic}[scale=0.35]{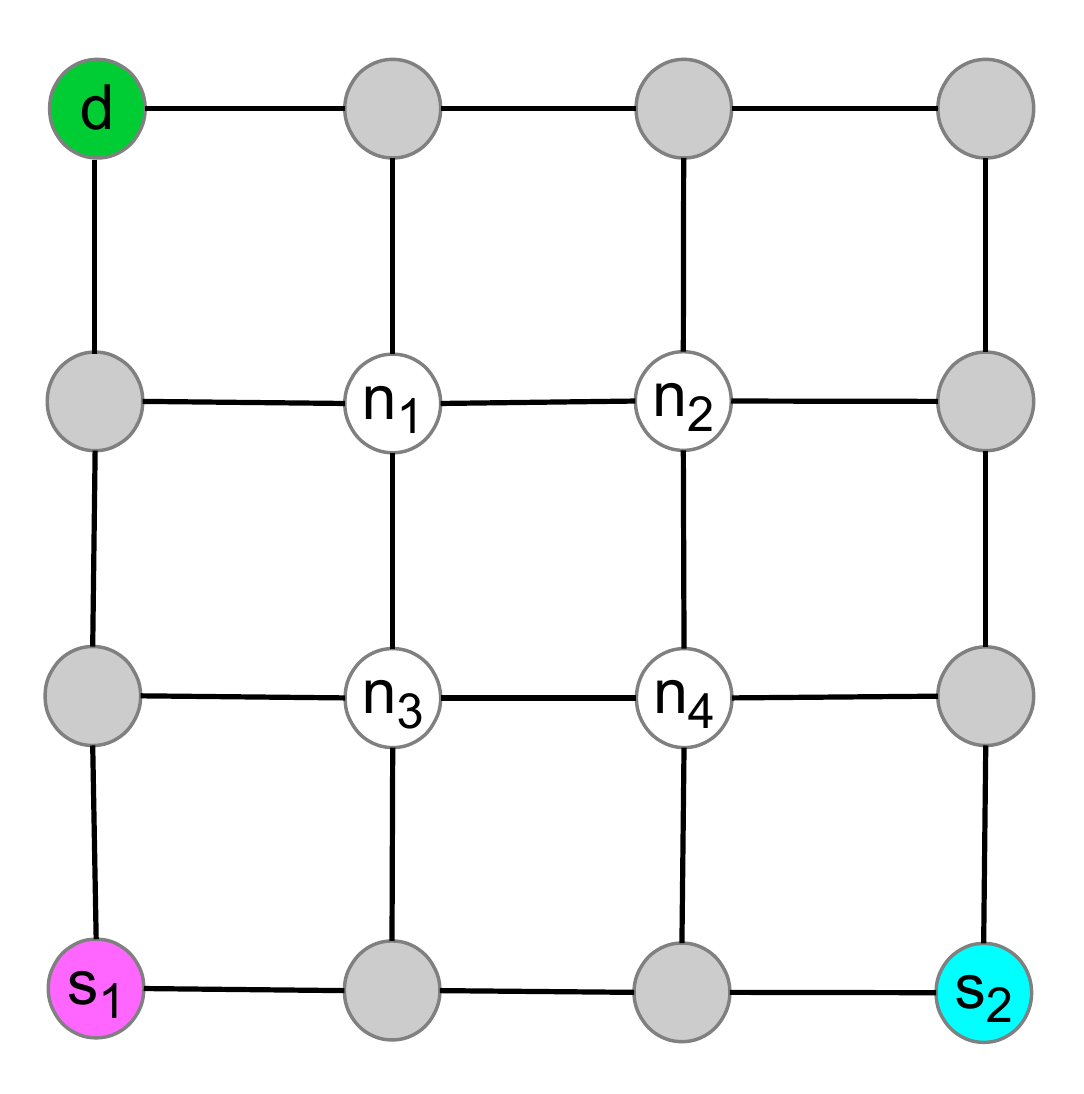}
			\put(44,-6){\small (a)}
		\end{overpic}
		\hspace{0.16in}
		\begin{overpic}[scale=0.25]{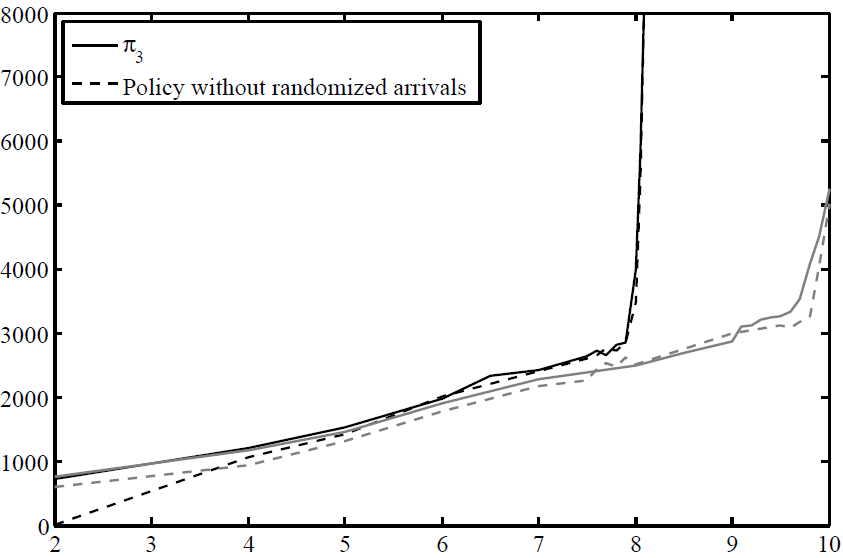} 
			\put(80,24){\vector(1,-2){4}}
			\put(83,12){\footnotesize $C=3$}
			\put(76,40){\vector(2,1){8}}
			\put(83,45){\footnotesize $C=2$}
			\put(22,-6){\small (b) ~~~Query Rate $\lambda$}
			\put(-6,10){\footnotesize \rotatebox{90}{Average Total Backlog}}
		\end{overpic}
		\hspace{0.06in}
		\hspace{0.16in}
		\begin{overpic}[scale=0.22]{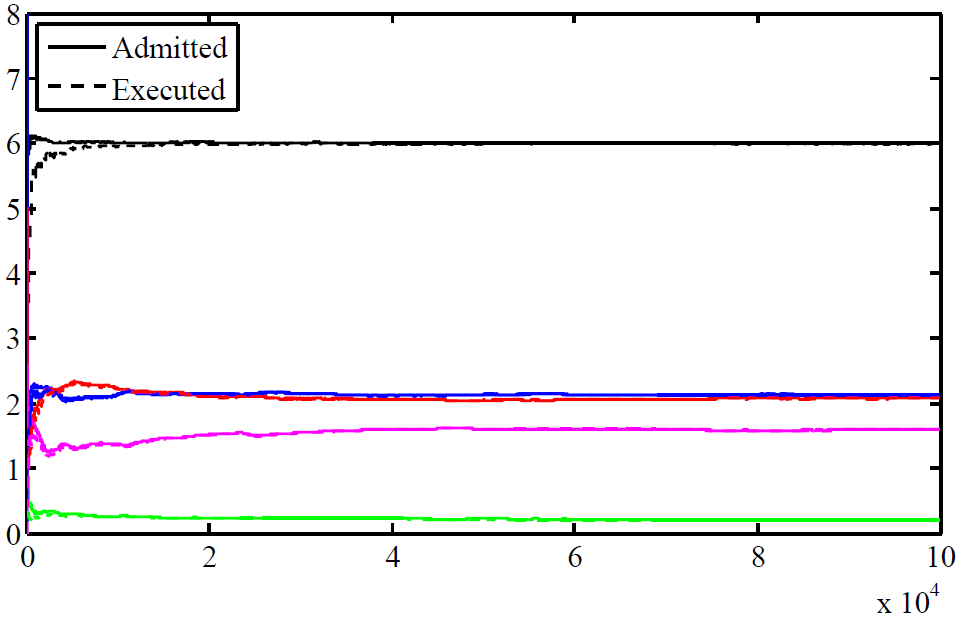}
			\put(10,-6){\small (c) ~~~~Time Slot Index}
			\put(-6,20){\footnotesize \rotatebox{90}{Running Average}}
			\put(60,49){\vector(1,-2){4}}
			\put(62,37){\footnotesize total}
			\put(24,23){\vector(1,2){4}}
			\put(27,33){\footnotesize node $n_1\&n_2$}
			\put(20,19){\vector(-1,2){6}}
			\put(5,32){\footnotesize node $n_4$}
			\put(59,10){\vector(1, 2){4}}
			\put(64,17){\footnotesize node $n_3$}
			
		\end{overpic}
		\caption{(a) The grid topology used for simulations. (b) Average total queue lengths for $\pi_3$ (solid lines), $\bar{\pi}_3$(dashed lines) vs. query rate for $C=2,C=3$.(c) Running averages of the query rate, allocations among embeddings  and total computations made for a simulation run of $\pi_3$ with $C=2,\lambda=6$.}
		\vspace{-0.3cm}
		\label{fig:grid}
		
	\end{figure*}
	
	The main result is that the policy above satisfies almost every query demand rate below $\lambda^*$, according to the choice of the control parameter $\epsilon_B$:
	\begin{theorem}\label{th:optimality_general_multiple}
		Policy $\pi_3$ stabilizes the network with multiple computing nodes for any query rate $\lambda < \left(1-\frac{\epsilon_B}{1+\epsilon_B}\right)\lambda^*$.
	\end{theorem}
	Next we provide a sketch of the proof of Theorem~\ref{th:optimality_general_multiple}.
	By exploiting the randomization that decouples routing from computation, we use classical Lyapunov drift techniques \cite{Neely05}, \cite[Theorem 4.5]{Georgiadis06} to prove strong stability of network queues, excluding the ones that take part in the computation. In particular we obtain the following bound in the Lyapunov drift expression:
	\begin{align*}
		& \Delta L(\mathbf{Q}(t), \mathbf{H}(t))\leq B - \sum_{n\in\mathcal{N_C}}C_nH_n(t)\label{eq:drift_multiple_2}+\sum_{n\in\mathcal{N}_C}\EE\left\{\tilde{A}^{(n)}(t)\right\}\\
		&\times\left((1+\epsilon_B)Q_n^{(0,n)}(t)+\sum_{i=1,2}Q_i^{(i,n)}(t)+H_n(t)\right) \notag \\
		&-\hspace{-0.15in}\sum_{\substack{(m,k)\in\mathcal{E}\\n\in\mathcal{N}_C \\i=0,1,2}}\hspace{-0.04in}\left(Q_m^{(i,n)}(t)-Q_k^{(i,n)}(t)\right)\left(\EE\left\{U_{mk}^{(i,n)}(t)-U_{km}^{(i,n)}(t)\right\}\right)\notag
	\end{align*}
	where we observe that the right-hand side above is minimized by the load balancing, routing and scheduling actions of our policy $\pi_3$. Combining the expressions above with an optimal randomized routing policy, we have
	\begin{lemma}\label{le:stablity_lemma_general_multiple}
		Under policy $\pi_3$, all queues $Q_{k}^{(i,n)}(t)$, $H_n(t)$, $\forall k, i, n$ are strongly stable for any $\lambda<\left(1-\frac{\epsilon_B}{1+\epsilon_B}\right)\lambda^*$.
	\end{lemma}
	To complete the proof of Theorem \ref{th:optimality_general_multiple} it remains to show strong stability for queues $X_n^{(i)}(t), Y_n(t)$. For this we extend the methodology of the previous section in a straightforward manner, e.g.~Lemmas \ref{le:threshold_based_general}, \ref{le:dominance_general}.
	
	\subsection{Discussion on $\pi_3$}
	
	The proposed policy $\pi_3$ is adaptive, it requires only local information for routing, and it can react to changes in the environment, providing robust efficient network computations. Using prior work \cite{Georgiadis06}, we may extend $\pi_3$ to the wireless case, covering applications within the range of edge and fog computing. Also note that the thresholds are used only for the proofs and are not necessary in the implementations of policies.
	
	Load balancing queries on different computation nodes requires some coordination, since there an agreement needed to be made and communicated between remote sources on the exact computation node that each query is using. This can be achieved by using an information exchange mechanism like \cite{Thaler98}. To facilitate timely coordination, it is possible to modify load balancing in the following way. Sources agree on a weighted round robin policy on how queries are assigned to computation nodes, and then frequently update weights in a coordinated fashion in order to balance the terms  in $\pi_3$.
	
	The main idea of the algorithms used is the intermediate queues $\mathcal{Y}_n(t)$ whose services are independent of the queues and routing controls in the network. This decouples the problem in a computation one (on which packets to be combined at the nodes), a "load balancing" one (on which embedding to use for new requests) and a communication (on routing and scheduling at links). The cost is using dummy packets in the network and being able to satisfy slightly smaller request rate than the maximum. A policy $\bar{\pi}_3$ that does not use the intermediate queues and does any computations possible at every slot is conjectured to achieve any $\lambda<\lambda^*$.
	
	\section{Simulation Results} \label{sec:eval}
	
	In order to illustrate the computational capacity and network behaviour under our algorithms, we examine a $4\times 4$ grid topology with four computation nodes as shown in Fig. \ref{fig:grid}(a).
	
	Each edge has a capacity of $R=5$ packets per slot and the computation nodes have the same capacity, $C$. The average total queue lengths versus the incoming query rate for policies $\pi_3$ with $\epsilon_B=0.01$ and a policy $\bar{pi}_3$ that does not use the randomized inputs are plotted in Fig. \ref{fig:grid} (b)  for $C=2, C=3$. We can observe that two policies achieve the same computational capacity, supporting our conjecture that the randomized inputs to $\mathcal{Q}_n^{(0,n)}$ are not really needed. In addition, $\bar{\pi}_3$ has fewer packets in the network in light loads. We can also note that when $C=2$, the computational capacity of the network is $\lambda^*=8$, therefore limited by the capacity of the computing nodes, while for $C=3$  it around $9.8$ and is therefore limited by the communication capacity of the network.
	
	Finally, Fig. \ref{fig:grid}(c) shows the empirical averages of the queries arriving to the system, computations allocated per computing node (i.e. on the load balancingphase of the algorithm) and computations executed. For this simulation, $C=2$ and $\lambda=6$, so it is an achievable computation rate by the network. We can observe that as time passes, the average computations made in the network matches the average query demand.          
	
	\section{Conclusions} \label{sec:conclusion}
	
	This paper is a first step towards understanding the performance limits of the interaction of big data and the underlying network resources. The big data challenge materializes as one of dealing with so-called 5 V’s (volume, variety, velocity, variability, complexity). We attempt to deep-dive into two of the five “V”’s, namely volume and velocity of data that stems from query streams. Our goal is to characterize the fundamental limits of the volume of queries that can be processed in the presence of limited resources 
	in a network setting. The study also aims to provide an understanding of the “velocity” dimension above i.e. how fast the generated volume of data can be processed. 
	
	There exist several directions for future work. A non-trivial extension of our work includes the case of computation queries with multiple possible computation graphs (DAGs) to choose from, where each computation graph may have several embeddings in the network graph. Another interesting repercussion is the scenario where the computation graph involves several types of operations, some of which are harder to perform than others. In that case, the rate at which computations are performed at computation nodes would depend on the type of operation to be performed.
	
	\bibliographystyle{IEEEtran}
	
	\bibliography{IEEEabrv,Computations_references_short}

\begin{thebibliography}{10}
\providecommand{\url}[1]{#1}
\csname url@samestyle\endcsname
\providecommand{\newblock}{\relax}
\providecommand{\bibinfo}[2]{#2}
\providecommand{\BIBentrySTDinterwordspacing}{\spaceskip=0pt\relax}
\providecommand{\BIBentryALTinterwordstretchfactor}{4}
\providecommand{\BIBentryALTinterwordspacing}{\spaceskip=\fontdimen2\font plus
\BIBentryALTinterwordstretchfactor\fontdimen3\font minus
  \fontdimen4\font\relax}
\providecommand{\BIBforeignlanguage}[2]{{%
\expandafter\ifx\csname l@#1\endcsname\relax
\typeout{** WARNING: IEEEtran.bst: No hyphenation pattern has been}%
\typeout{** loaded for the language `#1'. Using the pattern for}%
\typeout{** the default language instead.}%
\else
\language=\csname l@#1\endcsname
\fi
#2}}
\providecommand{\BIBdecl}{\relax}
\BIBdecl

\bibitem{Bonomi12}
F.~Bonomi, R.~Milito, J.~Zhu, and S.~Addepalli, ``Fog computing and its role in
  the internet of things,'' in \emph{MCC}, 2012.

\bibitem{Appuswami11}
R.~Appuswamy, M.~Franceschetti, N.~Karamchandani, and K.~Zeger, ``Network
  coding for computing: Cut-set bounds,'' \emph{IEEE Trans. Inf. Theory}, pp.
  1015--1030, 2011.

\bibitem{Shah11}
V.~Shah, B.~Dey, and D.~Manjunath, ``Network flows for functions,'' in
  \emph{ISIT}, 2011.

\bibitem{Liu13}
J.~Liu, C.~H. Xia, N.~B. Shroff, and X.~Zhang, ``On distributed computation
  rate optimization for deploying cloud computing programming frameworks,''
  \emph{ACM SIGMETRICS}, 2013.

\bibitem{Kannan13}
S.~Kannan and P.~Viswanath, ``Multi-session function computation and
  multicasting in undirected graphs,'' \emph{IEEE J. Sel. Areas. Commun.}, pp.
  702--713, 2013.

\bibitem{Giridhar05}
A.~Giridhar and P.~Kumar, ``Computing and communicating functions over sensor
  networks,'' \emph{IEEE J. Sel. Areas. Commun.}, pp. 755--764, 2005.

\bibitem{Banerjee12}
S.~Banerjee, P.~Gupta, and S.~Shakkottai, ``Towards a queueing-based framework
  for in-network function computation,'' \emph{Queueing Syst.}, pp. 219--250,
  2012.

\bibitem{Massoulie07}
L.~Massoulie, A.~Twigg, C.~Gkantsidis, and P.~Rodriguez, ``Randomized
  decentralized broadcasting algorithms,'' in \emph{IEEE INFOCOM}, 2007.

\bibitem{Zhao10}
H.~C. Zhao, C.~H. Xia, Z.~Liu, and D.~Towsley, ``A unified modeling framework
  for distributed resource allocation of general fork and join processing
  networks,'' in \emph{ACM SIGMETRICS}, 2010.

\bibitem{Jiang09}
L.~Jiang and J.~Walrand, ``Stable and utility-maximizing scheduling for
  stochastic processing networks,'' in \emph{Allerton}, 2009.

\bibitem{Huang11}
L.~Huang and M.~J. Neely, ``Utility optimal scheduling in processing
  networks,'' \emph{Perform. Evaluation}, pp. 1002--1021, 2011.

\bibitem{Vyavahare14}
P.~Vyavahare, N.~Limaye, and D.~Manjunath, ``Optimal embedding of functions for
  in-network computation: Complexity analysis and algorithms,'' \emph{CoRR},
  vol. abs/1401.2518, 2014.

\bibitem{Georgiadis06}
M.~J.~N. L.~Georgiadis and L.~Tassiulas, ``Resource allocation and cross-layer
  control in wireless networks,'' \emph{Foundations and Trends® in
  Networking}, vol.~1, 2006.

\bibitem{Tassulas92}
L.~Tassiulas and A.~Ephremides, ``Stability properties of constrained queueing
  systems and scheduling policies for maximum throughput in multihop radio
  networks,'' \emph{IEEE Trans. Autom. Control}, pp. 1936--1948, 1992.

\bibitem{Neely05}
M.~Neely, E.~Modiano, and C.~Rohrs, ``Dynamic power allocation and routing for
  time-varying wireless networks,'' \emph{IEEE J. Sel. Areas. Commun.}, pp.
  89--103, 2005.

\bibitem{Humes94}
J.~Humes, C., ``A regulator stabilization technique: Kumar-seidman revisited,''
  \emph{Automatic Control, IEEE Transactions on}, pp. 191--196, 1994.

\bibitem{Wu07}
X.~Wu, R.~Srikant, and J.~Perkins, ``Scheduling efficiency of distributed
  greedy scheduling algorithms in wireless networks,'' \emph{Mobile Computing,
  IEEE Transactions on}, pp. 595--605, 2007.

\bibitem{Joo09}
C.~Joo, X.~Lin, and N.~Shroff, ``Greedy maximal matching: Performance limits
  for arbitrary network graphs under the node-exclusive interference model,''
  \emph{Automatic Control, IEEE Transactions on}, pp. 2734--2744, 2009.

\bibitem{Thaler98}
D.~G. Thaler and C.~V. Ravishankar, ``Using name-based mappings to increase hit
  rates,'' \emph{IEEE/ACM Trans. Netw.}, pp. 1--14, 1998.

\bibitem{Neely10}
M.~J. Neely, ``Stochastic network optimization with application to
  communication and queueing systems,'' \emph{Synthesis Lectures on
  Communication Networks}, 2010.

\end{thebibliography}

\appendix



\subsection{Proof of Lemma 
	\ref{le:threshold_based_disjoint}}

All queues in $\mathcal{G}_1$ are strongly stable for $\lambda<\lambda^*$, since backpressure routing is applied and $n$ is the destination node, hence to prove the result we study the stability of the queues in node $n$.		      

As explained above, whenever condition ~(\ref{eq:condition_Cn}) is true, there are $C_n$ computations done in node $n$, therefore $C_n$ packets are removed from both queues. 

Now we turn to the choice of $\overline{X}$. The idea is to choose a threshold big enough for~(\ref{eq:condition_Cn}) to be satisfied with high probability if $X_n^{(1)}(t)+X_n^{(2)}(t)\geq 2C_n + \overline{X}$. For any $\lambda<\lambda^*$, all queues in $\mathcal{N}_1$ are strongly stable since backpressure is applied at this network, therefore steady state stable as well \cite{Neely10}. Therefore we have
\begin{align}
	\lim\limits_{t\rightarrow\infty}\PP\{X(t)>x\}=p(x)\\ \nonumber
	\text{and} \\
	\lim\limits_{x\rightarrow\infty}\lim\limits_{t\rightarrow\infty}\PP\{X(t)>x\}=0
	,\end{align}
from which we can deduce that, for every $\epsilon>0$, there is a $\overline{X}=\overline{X}(\epsilon)$ such that 
\begin{equation}\label{eq:bound_network_pkts}
	p(x)<\epsilon, \forall x>\overline{X}
	.\end{equation}
For any $\epsilon$, we can then find a $\overline{X}=\overline{X}(\epsilon)$ such that \eqref{eq:bound_network_pkts} is true. By construction of $\pi_1'$, the average service for each queue $X_n^{(i)}(t)$ is zero if $X_n^{(1)}(t)+X_n^{(2)}(t)< 2C_n + \overline{X}$. On the other hand, we can bound $\mathbb{E}\left\{Z(t)\right\}$ for $X_n^{(1)}(t)+X_n^{(2)}(t) \geq 2C_n + \overline{X}(\epsilon)$ using the following reasoning:  Define $Z'(t)$ such that 
\begin{equation}\nonumber
	Z'(t) = \begin{cases}
		C_n, \text{if } X_n^{(1)}(t)+X_n^{(2)}(t)\geq 2C_n + \overline{X} \text{ and } \eqref{eq:condition_Cn} \text{ is true}\\
		0, \text{otherwise} 
	\end{cases} 
\end{equation}
Clearly, $Z(t)= Z'(t)$ if \eqref{eq:condition_Cn} is true and if $X_n^{(1)}(t)+X_n^{(2)}(t)< 2C_n + \overline{X}$ (in the latter case both are zero) and $Z(t)\geq Z'(t)$ if $X_n^{(1)}(t)+X_n^{(2)}(t)\geq 2C_n + \overline{X}$ but \eqref{eq:condition_Cn} is not true (there may still be pairs in the computation node). The latter event happens with probability $\PP\left(X(t) >\overline{X}(\epsilon)\right) = p(\overline{X}(\epsilon)) < \epsilon$ so we have, for that
\begin{align}
	\EE\left\{Z(t)\right\}\geq\EE\left\{Z'(t)\right\}=(1-\epsilon)C_n, \\ \nonumber \text{for }X_n^{(1)}(t)+X_n^{(2)}(t)\geq 2C_n + \overline{X}
	.\end{align} The probability distribution $p(x)$ depends on the network $\mathcal{G}_1$  and the query arrival rate (since it depends on the probability distribution of $\left({Q}_k^{i}(t)\right)_{k\in\mathcal{N}_1, i\in\left\{1,2\right\}}$, hence the dependence of $\overline{X}$ on $\mathcal{G}_1, \lambda$.  

Now define $\delta=(\lambda^*-\lambda)/2$ and select a $\overline{X}$ such that (\ref{eq:bound_network_pkts}) holds for some $\epsilon < 1 - (\lambda+\delta)/C_n$. Define $L(\mathbf{q}) = \frac{1}{2}(q_1+q_2)$ and the $T$-slot drift $\Delta_T(\mathbf{q})=\EE\left\{L(\mathbf{X}_n(t+T))-L(\mathbf{X}_n(t))|\mathbf{Q}_n(t)=\mathbf{q}\right\}$. In addition, we define $A_i'(t)=\sum_{k\in\mathcal{N}(n)}U_{kn}^{(i)}(t)$ the arrivals at queue $X^{(i)}(t)$. Note that $A_i'(t)$ is a function of the queue state of the network $\mathcal{G}_1$ only. Since this network is strongly stable, for any $\delta>0$, there exists a $T_0<\infty$ such that 
\begin{equation}\nonumber
	\left|\frac{1}{T}\sum\limits_{\tau=t}^{t+T-1}\EE\{A_i'(\tau)|\mathbf{Q}(t)=\mathbf{q}\}-\lambda\right|<\delta ,\quad \forall i\in\{1,2\}\\
\end{equation}
for all $T>T_0, \mathbf{q}\in\mathbb{Z}_+^{2N_1}$. Choosing such a $T$, we have
\begin{equation} \nonumber
	\Delta_T(\mathbf{X}_n(t)) \leq DT^2 +\sum_{i=1}^2X_n^{(i)}(t)\sum\limits_{\tau=t}^{t+T}\left(\EE\{A'_i(\tau)\}-\EE\{Z(\tau)\}\right)
\end{equation}  
where $D=\frac{1}{2}\left(C_n^2 + (\sum_{j\in\mathcal{N}_{in}(n)}R_{jn})^2\right)$.  The $T-$slot drift is bounded by a constant for $X_n^{(1)}(t)+X_n^{(2)}(t)\leq2TC_n+\overline{X}$, while for $X_n^{(1)}(t)+X_n^{(2)}(t)>2TC_n+\overline{X}$ we can argue that there are at least $TC_n$ pairs in the computation node at the beginning of slot $t$ and the condition of $\pi_1'$ holds for every of the next $T$ slots therefore
\begin{equation}\nonumber
	\Delta_T(\mathbf{X}_n(t)) \leq DT^2 - (X_n^{(1)}(t)+X_n^{(2)}(t))T((1-\epsilon)C_n-\left(\lambda+\delta)\right) 
\end{equation}
From the choices of the threshold, we can deduce that there exists some $\delta'>0$ such that $\Delta_T(\mathbf{X}_n(t)) \leq DT^2 - T\delta'(X_n^{(1)}(t)+X_n^{(1)}(t))$ for $X_n^{(1)}(t)+X_n^{(1)}(t)>2TC_n+\overline{X}$, therefore $\mathcal{X}_n^{(1)}, \mathcal{X}_n^{(2)}$ are strongly stable. The dependence of the threshold $\overline{X}$ on $\lambda,C_n$ comes from the choice of $\epsilon$ in the proof. 

What is left is to prove stability of all queues in $\mathcal{G}_2$. Indeed, since $\mathcal{X}_n^{(1)}, \mathcal{X}_n^{(2)}$ are strongly stable, the arrival process at this network is a Markov modulated process with mean rate $\lambda$; this network is then stable as $\lambda<\lambda^*$ and backpressure routing is applied.

\subsection{Proof of Lemma \ref{le:stablity_lemma_general_multiple}}
The proof is similar to the one in \cite{Neely05}. Consider any query rate $\lambda<\left(1-\frac{\epsilon_B}{1+\epsilon_B}\right)\lambda^* = \frac{\lambda^*}{1+\epsilon_B}:=\bar{\lambda}$ and define $\delta = \bar{\lambda} - \lambda$. Also, define the set of commodities $\mathcal{C}=\left\{1\rightarrow n, 2\rightarrow n, n\rightarrow d, \forall n\in\mathcal{N}_C \right\}$. Then, the vector $\left({\lambda}_n^*\right)_{n\in\mathcal{N}_C}$ corresponding to the vector satisfying the conditions of Threorem \ref{th:upperbound2} is a point on the boundary of the throughput region of the network, therefore the traffic vector from which $\frac{\lambda^*_n}{1+\epsilon_B}$ is in the interior of the throughput region (the corresponding traffic demand vector of the multicommodity problem is strictly dominated by the one that is defined by  $\left(\lambda^*_n\right)_n$). Now, for the query rate $\lambda$, we define a traffic demand vector such that the demand rate for commodities $1\rightarrow n, 2\rightarrow n$ is $\lambda_n = \bar{\lambda}_n-\delta/N_C$ and for commodity $n\rightarrow d$ is $(1+\epsilon_B)\lambda_n = (1+\epsilon_B)(\bar{\lambda}_n-\delta/N_C)=\lambda^*_n - (1+\epsilon_B)\delta/N_C$. We can see that this vector is also strictly dominated by the one defined by $\left(\lambda^*_n\right)_n$, therefore it lies inside the throughput region of the multicommodity flow problem, which in turn implies that there exist a flow allocation $\left(\bar{f}_{kl}^{(i,n)}\right)$ such that the constraints are met with inequality (see e.g. the proof of \cite[Theorem 4.5]{Georgiadis06}); define then $\epsilon'$ the munimum difference of the RHS minus the LHS over all flow conservation inequality constraints.  

A corresponding policy that achieves satisfies the query rate $\lambda$ is then $\bar{\pi}$ according to which $U_{km}^{(i,n)}(t)=\bar{f}_{km}^{(i,n)}$ for all slots $t$ and $n^*(t)=n$ with probability $\frac{\bar{\lambda}_n-\delta/N_C}{\lambda}$. However, from eq.~(\ref{eq:drift_multiple_2}) is follows that policy $\pi_3$ minimizes at every slot the bound on the drift of the Lyapunov function, therefore we have 
\begin{align}\nonumber
	&\Delta L(\mathbf{Q}(t), \mathbf{H}(t))\leq B-\sum_{n\in\mathcal{N}_C}H_n(t)\left(C_n-\lambda_n\right)\\ \nonumber
	&-\sum_{n\in\mathcal{N}_C}\bigg[Q_n^{(0,n)}(t)\left(\sum_{l\in\mathcal{N}(n)}\bar{f}_{nl}^{(0,n)}-\bar{f}_{ln}^{(0,n)}-(1+\epsilon_B){\lambda}_{n}\right) \\ \nonumber
	& - \sum_{i=1}^{2}Q_i^{(i,n)}(t)\left(\sum_{l\in\mathcal{N}(n)}(\bar{f}_{il}^{(i,n)}-\bar{f}_{li}^{(i,n)})-{\lambda}_n\right)\\ \nonumber
	& - \sum_{k\in\mathcal{V}}\sum_{i=1,i\neq k}^{2}Q_k^{(i,n)}(t)\left(\sum_{l\in\mathcal{N}(n)}\left(\bar{f}_{kl}^{(i,n)}-\bar{f}_{lk}^{(i,n)}(t)\right)\right)\\ \label{eq:drift_multiple_sss}
	& - \sum_{k\in\mathcal{N}}Q_k^{(0,n)}(t)\left(\bar{f}_{kl}^{(0,n)}-\bar{f}_{lk}^{(0,n)}\right) \bigg]\\ \nonumber
	&\leq B - \epsilon'\left(\sum_{k\in\mathcal{V}, i\in\{0,1,2\}, n\in\mathcal{N}_C}Q_k^{(i,n)}(t)  + \sum_{n\in\mathcal{N}_C}H_n(t)\right),
\end{align}     
Inequality (\ref{eq:drift_multiple_sss}) follows from replacing the actions taken by policy $\bar{\pi}$ and the fact that $\pi_3$ minimizes the drift.The latter expression also implies that the queues are strongly stable, completing the proof. 


\subsection{Proof of Theorem \ref{th:optimality_general_multiple}}

Theorem  \ref{th:optimality_general_multiple} is a direct consequence of Lemma \ref{le:stablity_lemma_general_multiple} and Lemmas \ref{le:threshold_based_general_multiple}, \ref{le:dominance_general_multiple} that follow. 

\begin{lemma}\label{le:threshold_based_general_multiple}
	There exists a threshold $\overline{X}$ such that the queues $X_n^{(i)}(t), Y_n(t)$ are stable under $\pi_3'$ for any $\lambda<\lambda^*\left(1 - \frac{\epsilon_B}{1+\epsilon_B}\right)$.
\end{lemma} 

The above lemma is proven using strong stability of $\mathbf{Q}(t), \mathbf{H}(t)]$, and its proof is given in the next subsection. 

\begin{lemma}\label{le:dominance_general_multiple}
	$X_n^{(i), \pi_3}(t) \leq_{\text{st}}X_n^{(i), \pi_3'}(t), \forall i\in\{1,2\}$ for all thresholds $\overline{X}$.
\end{lemma}
The proof of the above Lemma is identical to the one for Lemma \ref{le:dominance_disjoint}. 

\subsection{Proof of Lemma 
	\ref{le:threshold_based_general_multiple}}

The proof for the queues $X_n^{(i)}(t)$ is similar  to the proof of Lemma~\ref{le:threshold_based_disjoint}. We define $\mathbf{S}(t)=\left[\mathbf{Q}(t) , \mathbf{H}(t)\right]$. Take any $\lambda<\left(1-\frac{\epsilon_B}{1+\epsilon_B}\right)\lambda^*$ and denote  $X_n(t)=\sum_{i=1}^{2}\sum_{m\in\mathcal{V}\setminus\{n\}}Q_m^{(i,n)}(t)$ the number of uncombined packets in the network that have to be combined at node $n$. Using the same arguments  of the proof of Lemma~\ref{le:threshold_based_disjoint} for each computing node $n\mathcal{N}_C$, we can argue that $C_n$ computations are made if
\begin{equation}
	X_n^{(1)}(t) + X_n^{(2)}(t)\geq 2C_n + X_n(t)
	.\end{equation}
However, policy $\pi_3'$ injects packets in the network and routes them in the same way as $\pi_3$ does, therefore the queues in $\mathbf{S}(t)$ are strongly stable for $\lambda<\left(1-\frac{\epsilon_B}{1+\epsilon_B}\right)\lambda^*$, which  implies that this process is positive recurrent with finite mean for every queue. A stationary probability distribution $p(\mathbf{s}), \mathbf{s}$ then exists, and $|p_t(\mathbf{s})-p(\mathbf{s})|_{TV}\rightarrow 0$. Since the arrivals $A^{(n)'}_i(t)=\sum_{k\in\mathcal{N}(n)}U_{kn}^{(i,n)}(t)$ at each queue $X_n^{(i)}(t)$ depend on $\mathbf{S}(t)$, it follows that for any $\delta>0$, there exists $T_0(\delta)>0$ such that
\begin{align}
	\left|\frac{1}{T}\sum\limits_{\tau=t}^{t+T-1}\EE\{A_i^{(n)'}(\tau)|\mathbf{S}(t)=\mathbf{s}\}-\bar{\lambda}_n\right|<\delta ,\quad \forall i\in\{1,2\}, \forall n\in\mathcal{N}_C\\
	\left|\frac{1}{T}\sum\limits_{\tau=t}^{t+T-1}\PP\{X_n(t)>x|\mathbf{S}(t)=\mathbf{s}\}-p^{(n)}(x)\right|<\delta, \forall x>0,\forall n\in\mathcal{N}_C 
\end{align}
for any $T>T_0(\delta)$ and $\mathbf{q}\in\mathbb{Z}_+^{N_CK}$. The quantities $\bar{\lambda}_n$ satisfy the conditions of Theorem \ref{th:upperbound2}. In addition, for every $\epsilon>0$ there is a $x_\epsilon$ such that $p^{(n)}(x)<\epsilon, \forall x>x_{\epsilon}$.

We now denote $\delta_n=C_n-\bar{\lambda}_n>0$ and choose $\delta=\min\limits_{n}\left[\frac{\delta_n}{4C_n}\right]$, some $T>T_0(\delta)$ and $\epsilon < \delta$. For the $T-$slot drift of the quadratic Lyapunov function at every node $n\in\mathcal{N}_C$ we have:
\begin{align}
	&\Delta^{(n)}_T(\mathbf{X}(t)) = \EE\left\{L(\mathbf{X}(t+T))-L(\mathbf{X}(t))|\mathbf{X}(t)\right\}\\
	&\leq D_nT^2 + \sum_{i=1}^2X_n^{(i)}(t)\sum_{\tau=t}^{t+T-1}\EE\left\{A_i^{(n)'}(t)-Z_n(t)|\mathbf{X}(t)\right\}
	.\end{align}  
The sum over $t$ can be written as follows for $\sum_{i}X_n^{(i)}(t)>2TC_n+\bar{X}$. 
\begin{align}
	&	\sum_{\tau=t}^{t+T-1}\EE\left\{A_i^{(n)'}(t)-Z_n(t)|\mathbf{X}(t)\right\} =\\ 	&\sum_{\tau=t}^{t+T-1}\EE\left\{A_i^{(n)'}(t)|\mathbf{X}(t)\right\} - 	\sum_{\tau=t}^{t+T-1}\EE\left\{Z_n(t)|\mathbf{X}(t)\right\} \leq \\
	&\sum_{\tau=t}^{t+T-1}\EE\left\{A_i^{(n)'}(t)|\mathbf{X}(t)\right\} -\sum_{\tau=t}^{t+T-1}\EE\left\{C_n\ind_{X_n(t)<\bar{X}}|\mathbf{X}(t)\right\}=\\
	& \sum_{\tau=t}^{t+T-1}\EE\left\{A_i^{(n)'}(t)|\mathbf{X}(t)\right\} -\sum_{\tau=t}^{t+T-1}C_n(1-\PP\left\{X_n(t)>\bar{X}|\mathbf{X}(t)\right\})\\
	&\leq T(\bar{\lambda}_n+\delta-C_n(1-p^{(n)}(\bar{X})-\delta))\\ \nonumber
	&\leq T\left(C_n - \delta_n+\delta -C_n(1-\epsilon - \delta)\right) \\ 
	&<T\left(2\frac{\delta}{4}+\frac{\delta}{4C_n}-\delta\right) \leq -\frac{\delta}{4} 
	.\end{align}
In the above, we have used the fact that since $\sum_{n,i}X_n^{(i)}(t)>2TC_n+\bar{X}$, at time $t+t'$ there are at least $(T-(t'-t))C_n$ pairs in node $n$. The first inequality is true since the right hand side assumes that computations are made only when $X_n(t)X\bar{X}$. For the last parts we used the definitions of the constants $\epsilon, \delta,\bar{X}$. For each computing node we then have 
\[
\Delta^{(n)}_T(\mathbf{X}(t)) \leq \begin{cases}
D'_n, \text{if } \sum_{i}X_n^{(i)}(t)\leq 2TC_n+\bar{X}\\
D_n - \frac{\delta}{4}T \sum_{i=1}^2X_n^{(i)}(t), \text{otherwise}
,\end{cases}
\]
where $D'_n = D_nT + 2(2TC_n+\bar{X})\sum_{k\in\mathcal{N}}R_{kn}<\infty$. The above can then be rewritten as 
\begin{equation}\label{eq:drifts_multiple_Tslot}
	\Delta^{(n)}_T(\mathbf{X}(t)) \leq \hat{D}_n - \frac{\delta}{4}T \sum_{i=1}^2X_n^{(i)}(t)
	,\end{equation}
where we have defined $\hat{D}_n = \max\left[D_n,D'n+2TC_n+\bar{X}\right]$. For the drift taking into account all queues with packets waiting to be combined $\Delta_n(\mathbf{X}(t))=\EE\left\{L(\mathbf{X}(t+T))-L(\mathbf{X}(t))\big| \mathbf{X}(t)\right\} = \sum_{n\in\mathcal{N}_C}\Delta^{(n)}_T(\mathbf{X}(t))$, it follows from \eqref{eq:drifts_multiple_Tslot} that, for $D^* = \sum_{n\in\mathcal{N}_C}\hat{D}_n$,  
\[
\Delta_T(\mathbf{X}(t)) \leq D^* - \frac{T\delta}{4}\sum_{n\in\mathcal{N}_C}\sum_{i=1}^{2}X_n^{(i)}(t)
,\]
thus $\mathbf{X}(t)$ are strongly stable. 
To finish with the proof, we should show that the queues $Y_n(t), n\in\mathcal{N}_C$ are strongly stable. Indeed, the arrival processes to these queues are
\begin{equation}
	A_Y^{(n)}(t) = \begin{cases}
		C_n, \text{if } \sum_{i}X_n^{(i)}(t)> 2C_n+\bar{X}\\
		0, \text{otherwise}
	\end{cases}
	,\end{equation}
thus the corresponding mean arrival rates are 
\begin{equation}
	\hat{\lambda}_n = C_n\limsup\limits_{T\rightarrow\infty}\frac{1}{T}\sum_{t=0}^{T-1}\mathbb{P}\left(\sum_{i=1}^{2}X_n^{(i)}(t)> 2C_n+\bar{X}\right)
	.\end{equation}
Strong stability of the queues $X_n^{(i)}(t)$ implies, in addition, that $\hat{\lambda}_n=\bar{\lambda}_n$. To see this, note that: 1) If $\hat{\lambda}_n<\bar{\lambda}_n$ then the queues $X_n^{(i)}(t)$ would be unstable and 2) No more pairs of packets than the ones received can be combined and send a packet to $Y_n(t)$.  For every $\delta>0$ then, since the queues $\mathbf{X}(t)$ are strongly stable, there exists $T_0(\delta)<\infty$ such that 
\begin{equation}\label{eq:sup_UB}
	\left|\frac{C_n}{T}\sum_{t=0}^{T-1}\PP\left(\sum_{i=1}^{2}X_n^{(i)}(t)> 2C_n+\bar{X}\right) - \bar{\lambda}_n\right| <\delta, \forall T>T_0(\delta)
\end{equation}

The average service rate of queue $Y_n(t)$ is, in turn, 
\begin{align}\nonumber
	\hat{\mu}_n&=\limsup\limits_{T\rightarrow\infty}\frac{1}{T}\sum_{t=0}^{T-1}\mathbb{E}\left\{\tilde{A}^{(n)}(t)(1 + B^{(n)}(t))\right\}\\ \nonumber
	&=\left(1+ {\epsilon_B }\right) \lim\limits_{T\rightarrow\infty}\frac{1}{T}\sum_{t=0}^{T-1}\mathbb{E}\left\{\tilde{A}^{(n)}(t)\right\} = (1+\epsilon_B)\bar{\lambda}_n 
	.\end{align}
The limit exists because it depends only on $\mathbf{Q}(t)$, which is a positive recurrent process with finite mean. Strong stability of $Y_n(t)$ can be shown exploiting the above discussion and (\ref{eq:sup_UB}) and taking an appropriate $T-$slot drift. 

\end{document}